\documentstyle[12pt,psfig]{article}
\begin{document}
\newcommand{\eos}{equation of state~}
\newcommand{\eoss}{equations of state~}
\newcommand{\eossp}{equations of state}
\newcommand{\eosp}{equation of state}
\newcommand{\Eos}{Equation of state~}
\newcommand{\Eosp}{Equation of state}
\newcommand{\Eoss}{Equations of state~}
\newcommand{\Eossp}{Equations of state}
\newcommand{\epb}{energy per baryon~}
\newcommand{\Epb}{Energy per baryon~}
\newcommand{\epbp}{energy per baryon}
\newcommand{\epbn}{energy per baryon number~}
\newcommand{\Epbn}{Energy per baryon number~}
\newcommand{\epbnp}{energy per baryon number}
\newcommand{\smh}{strange matter hypothesis~}
\newcommand{\smhp}{strange matter hypothesis}
\newcommand{\beqn}{\begin{eqnarray}}
\newcommand{\eeqn}{\end{eqnarray}}
\newcommand{\okgr}{\Omega_{\rm K}}
\newcommand{\ognm}{\Omega^\nu_m}
\newcommand{\ogtm}{\Omega^T_m}
\newcommand{\tsubk}{t_{\rm K}}
\newcommand{\pkgr}{P_{\rm K}}
\newcommand{\penm}{P^\nu_m}
\newcommand{\petm}{P^T_m}
\newcommand{\pet}{P^T}
\newcommand{\rmkm}{\rm km}
\newcommand{\msec}{\rm msec}
\newcommand{\gsim}{\stackrel{\textstyle >}{_\sim}}
\newcommand{\lsim}{\stackrel{\textstyle <}{_\sim}}
\newcommand{\fmmo}{{\rm fm}^{-1}}
\newcommand{\fmmt}{{\rm fm}^{-3}}
\newcommand{\mevt}{{\rm MeV/fm}^3}
\newcommand{\mev}{\rm MeV}
\newcommand{\mevfm}{{\rm MeV\;fm}}
\newcommand{\gcmt}{{\rm g/cm}^3}
\newcommand{\cmt}{{\rm cm}^3}
\newcommand{\gcmsq}{{{\rm g}\; {\rm cm}}^2}
\newcommand{\cmssm}{{{\rm cm}^2\,{\rm s}^{-1}}}
\newcommand{\rmhv}{{\rm HV}}
\newcommand{\rmhfv}{{\rm HFV}}
\newcommand{\rmbonn}{{\rm Bonn}}
\newcommand{\rmhea}{{\rm HEA}}
\newcommand{\rmbhv}{{\rm Bonn+HV}}
\newcommand{\rmhhfv}{{\rm HEA+HFV}}
\newcommand{\ebnhv}{{\Lambda^{00}_{\rm Bonn}}+{\rm HV}}
\newcommand{\ehehfv}{{\Lambda^{00}_{\rm HEA}}+{\rm HFV}}
\newcommand{\eBrBhfv}{{\Lambda^{\rm BHF}_{\rm BroB}}+{\rm HFV}}
\newcommand{\ebji}{BJ(I)~}
\newcommand{\epanc}{Pan(C)~}
\newcommand{\efp}{{\rm FP(V}_{14}+{\rm TNI)}}
\newcommand{\ewut}{{\rm WFF(UV}_{14}+{\rm TNI)}}
\newcommand{\ewuu}{{\rm WFF(UV}_{14}+{\rm UVII)}}
\newcommand{\ewau}{{\rm WFF(AV}_{14}+{\rm UVII)}}
\newcommand{\egva}{{\rm G}_{\rm V1}}
\newcommand{\egtw}{{\rm G}^\pi_{200}}
\newcommand{\egth}{{\rm G}_{300}}
\newcommand{\egthpi}{{\rm G}^\pi_{300}}
\newcommand{\egdcma}{{\rm G}^{\rm DCM1}_{\rm B180}}
\newcommand{\egdcmb}{{\rm G}^{\rm DCM1}_{225}}
\newcommand{\egdcmc}{{\rm G}^{\rm DCM2}_{\rm B180}}
\newcommand{\egdcmd}{{\rm G}^{\rm DCM2}_{\rm 265}}
\newcommand{\ecenm}{{\epsilon_c}/{\epsilon_0}}
\newcommand{\igcm}{{\rm log}\, I/{\rm g}\;{\rm cm}^2}
\newcommand{\zfw}{z_{\rm F}}
\newcommand{\zbw}{z_{\rm B}}
\newcommand{\zpole}{z_{\rm p}}
\newcommand{\bag}{B^{1/4}}
\newcommand{\edrip}{\epsilon_{\rm drip}}
\newcommand{\pdrip}{P_{\rm drip}}
\newcommand{\pbps}{P_{\rm BPS}}
\newcommand{\crust}{\rm crust}
\newcommand{\rcrust}{R_{\rm crust}}
\newcommand{\vcrust}{V_{\rm crust}}
\newcommand{\rgap}{R_{\rm gap}}
\newcommand{\mcrust}{M_{\rm crust}}
\newcommand{\mcore}{M_{\rm core}}
\newcommand{\rcore}{R_{\rm core}}
\newcommand{\ecrusti}{\epsilon_{\rm crust}}
\newcommand{\icrust}{I_{\rm crust}}
\newcommand{\itotal}{I_{\rm total}}
\newcommand{\rsurf}{R_{\rm surf}}
\newcommand{\rdrip}{R_{\rm drip}}
\newcommand{\eps}{\epsilon}
\newcommand{\epsc}{\epsilon_c}
\newcommand{\epsnm}{\epsilon_0}
\newcommand{\bfpm}{\mbox{\mbox{\boldmath$p$}}}
\newcommand{\bfxm}{\mbox{\mbox{\boldmath$x$}}}
\newcommand{\bcdot}{\mbox{\mbox{\boldmath$\cdot$}}}
\newcommand{\msun}{M_{\odot}}
\newcommand{\moip}{moment of inertia}
\newcommand{\moi}{moment of inertia~}
\newcounter{sctn}
\newcounter{subsctn}[sctn]
\newcommand{\sctn}[1]{~\\ \refstepcounter{sctn} {\bf \thesctn~~ #1} \\ }
\newcommand{\subsctn}[1]
{~\\ \refstepcounter{subsctn} {\bf \thesctn.\thesubsctn~~ #1}\\}

\newcommand{\dateofdoc}{\today}

\newcommand{\lbl}{\begin{flushright} LBNL--39268 \\[5ex] \end{flushright}}

\newcommand{\tit}{\bf Dense Stellar Matter and Structure of Neutron Stars}

\newcommand{\doe}
{This work was supported by the Director, Office of Energy Research,
Office of High Energy and Nuclear Physics, Division of Nuclear Physics, and 
of the U.S. Department of Energy under Contract DE-AC03-76SF00098.}

\newcommand{\adrlbl} {{Nuclear Science Division and \\ Institute for Nuclear \&
    Particle Astrophysics \\ Lawrence Berkeley National Laboratory \\ 
    University of California \\ Berkeley, California 94720, USA \\ Email:
    fweber@nsdssd.lbl.gov} \\[4ex]}

\newcommand{\adrlblb}{{Nuclear Science Division and \\ Institute for Nuclear \&
    Particle Astrophysics \\ Lawrence Berkeley National Laboratory \\ 
    University of California \\ Berkeley, California 94720, USA \\ Email:
    fweber@nsdssd.lbl.gov} \\[2ex]}

\newcommand{\adweb} {Home institute: Ludwig--Maximilians University of Munich,
  Institute for Theoretical Physics, Theresienstrasse 37, W-80333 Munich,
  Germany.}

\newcommand{\auth}{Fridolin Weber{\footnote{\adweb}} and 
Norman K. Glendenning \\[6ex]}

\newcommand{\authb}{Fridolin Weber and Norman K. Glendenning \\[4ex]}

\newcommand{\adr}{{Institute for Theoretical Physics \\
  Ludwig--Maximilians University of Munich \\ 
  Theresienstrasse 37, W-80333 Munich, Germany \\
  Email: fweber@mfl.sue.physik.uni-muenchen.de} \\[4ex]}

\newcommand{\adrb}{Institute for Theoretical Physics \\
  Ludwig--Maximilians University of Munich \\ 
  Theresienstrasse 37, W-80333 Munich, Germany \\
  Email: fweber@mfl.sue.physik.uni-muenchen.de \\[2ex]}

\newcommand{\adlbl}
{Nuclear Science Division, Lawrence Berkeley National Laboratory,
University of California, Berkeley, California 94720, USA.}

\begin{titlepage}
\lbl
\renewcommand{\thefootnote}{\fnsymbol{footnote}}
\setcounter{footnote}{0}
\begin{center}
\begin{Large}
\tit \\[7ex]
\end{Large}
\renewcommand{\thefootnote}{\fnsymbol{footnote}}
\begin{large}
\auth
\end{large}
\adrlbl
\dateofdoc \\[30ex]
\end{center}

\begin{quote}
\begin{center} 
{Presented by F. Weber at the \\
3rd Mario Sch\"onberg School on Physics \\
Jo${\tilde{\rm a}}$o Pessoa, Brazil \\
July 22 -- August 2, 1996 \\
To be published in the Brazilian Journal \\
of Teaching Physics}
\end{center}
\end{quote}
\end{titlepage}

\begin{titlepage}
\tableofcontents
\end{titlepage}
\renewcommand{\thefootnote}{\arabic{footnote}}
\setcounter{footnote}{0}
\newpage

\renewcommand{\thefootnote}{\arabic{footnote}}
\setcounter{footnote}{0}
\begin{center}
\begin{Large}
\tit \\[4ex]
\end{Large}
\begin{large}
\authb
\end{large}
\adrlblb
\end{center}
\vskip 1.0truecm

\begin{abstract}
  After giving an overview of the history and idea of neutron stars, I
  shall discuss, in part one of my lectures, the physics of dense
  neutron star matter. In this connection concepts like chemical
  equilibrium and electric charge neutrality will be outlined, and the
  baryonic and mesonic degrees of freedom of neutron star matter as
  well as the transition of confined hadronic matter into quark matter
  at supernuclear densities are studied. Special emphasis is put onto
  the possible absolute stability of strange quark matter.  Finally
  models for the equation of state of dense neutron star matter, which
  account for various physical possibilities concerning the unknown
  behavior of superdense neutron star matter, will be derived in the
  framework of relativistic nuclear field theory. Part two of my
  lectures deals with the construction of models of static as well as
  rapidly rotating neutron stars and an investigation of the cooling
  behavior of such objects. Both is performed in the framework of
  Einstein's theory of general relativity.  For this purpose the broad
  collection of models for the equation of state, derived in part one,
  will be used as an input. Finally, in part three a comparison of the
  theoretically computed neutron star properties (e.g., masses, radii,
  moments of inertia, red and blueshifts, limiting rotational periods,
  cooling behavior) with the body of observed data will be performed.
  From this conclusions about the interior structure of neutron stars
  -- and thus the behavior of matter at extreme densities -- will be
  drawn.
\end{abstract}

\section{Introduction}\label{sec:intro}

Neutron stars contain matter in one of the densest forms found in the
universe.  Matter in their cores possesses densities ranging from a
few times $\rho_0$ to an order of magnitude higher.  Here
$\rho_0=0.15$ nucleons/fm$^3$ denotes the density of normal nuclear
matter, which corresponds to a mass density of $2.5\times 10^{14}$
g/cm$^3$. The number of baryons forming a neutron star is of the order
of $A \approx 10^{57}$.  The understanding of matter under such
extreme conditions of density is one of the central but also most
complex problems of physics.

Neutron stars are associated with two classes of astrophysical
objects: Pulsars \cite{manchester77:a}, which are generally accepted
to be rotating neutron stars (the fastest so far observed ones have
rotational periods of $P=1.6$ ms, which corresponds to about 620
rotations per second), and compact X-ray sources (e.g., Her X-1 and
Vela X-1), certain of which are neutron stars in close binary orbits
with an ordinary star. The first millisecond pulsar was discovered in
1982 \cite{backer82:a}, and in the next seven years about one a year
has been found. The situation has changed radically with the recent
discovery of an anomalously large population of millisecond pulsars in
globular clusters \cite{manchester91:a}, where the density of stars is
roughly 1000 times that in the field of the galaxy and which are
therefore very favorable environments for the formation of rapidly
rotating pulsars that have been spun up by means of mass accretion
from a binary companion.  At present about 700 pulsars are known, and
the discovery rate of new ones is rather high.

As just outlined, neutron stars are objects of highly compressed
matter so that the geometry of space-time is changed considerably from
flat space.  Thus for the construction of realistic models of rapidly
rotating compact stars one has to resort to Einstein's theory of
general relativity
\cite{friedman86:a,friedman89:a,lattimer90:a,weber90:a,weber91:b,weber91:d}.
The \eos of the stellar matter, i.e., pressure as a function of energy
density, is the basic input quantity whose knowledge over a broad
range of densities (ranging from the density of iron at the star's
surface up to $\sim 15$ times the density of normal nuclear matter
reached in the cores of massive stars) is necessary in order to solve
Einstein's equations.  Unfortunately the physical behavior of matter
under such extreme densities as in the cores of massive stars is
rather uncertain and the associated \eos is only poorly known. The
models derived for it differ considerably with respect to the
dependence of pressure on density, which has its origin in various
sources. To mention several are: (1) the many-body technique used to
determine the \eosp, (2) the model for the nucleon-nucleon
interaction, (3) description of electrically charge neutral neutron
star matter in terms of either (a) only neutrons, (b) neutrons and
protons in $\beta$ equilibrium with electrons and muons, or (c)
nucleons, hyperons and more massive baryon states in $\beta$
equilibrium with leptons, (4) inclusion of pion/kaon condensation, and
(5) treatment of the transition of confined hadronic matter into quark
matter.  It is the purpose of this work to outline the present status
of dense matter calculations, explore the compatibility of the
properties of non-rotating and rotating compact star models, which are
constructed for a collection of \eoss which accounts for items
(1)--(5) from above, with observed data, and investigate the
properties of hypothetical strange (quark matter) stars
\cite{witten84:a,alcock86:a,haensel86:a}. The latter would constitute
the true nature of neutron stars if strange quark matter should be
more stable than ordinary nucleonic matter.

\section{\Eos of dense neutron star matter}\label{sec:neos}

\subsection{Theoretical framework}\label{ssec:theor}

\subsubsection{Non-relativistic approach}\label{sssec:nonr}

For non-relativistic models, the starting point is a phenomenological
nucleon-nucleon interaction. In the case of the \eoss
reported here, different two-nucleon potentials (denoted $V_{ij}$) 
which fit nucleon-nucleon scattering data and deuteron properties have 
been employed. Most of these two-nucleon potentials are supplemented with 
three-nucleon interactions (denoted $V_{ijk}$). The hamiltonian
is of the form
\beqn
         H \; = \; \sum_i \left( {{-\,\hbar^2}\over{2\, m}} \right)
              \, \nabla_i^2
            \; + \; \sum_{i<j} V_{ij} \; + \; \sum_{i<j<k} V_{ijk} \; .
\label{eq:hamil}
\eeqn
The many-body method adopted to solve the Schroedinger equation is based on 
the variational approach \cite{sprung72:a,day78:a,pandharipande79:a}
where a variational trial 
function $|\Psi_v>$ is constructed from a symmetrized product of two-body 
correlation operators ($F_{ij}$) acting on an unperturbed ground-state, i.e.,
\beqn
      |\Psi_v> \; = \; \left[ \hat S \; \prod_{i<j} \; F_{ij} \right] |\Phi>
      \; ,
\label{eq:psiv}
\eeqn
where $|\Phi>$ denotes the antisymmetrized Fermi-gas wave function,
\beqn
       |\Phi> \; = \; \hat A \; \prod_j \; 
     {\rm exp}(i\, \bfpm_j \bcdot \bfxm_j)
     \; .
\label{eq:phi} 
\eeqn
The correlation operator contains variational parameters which 
are varied to minimize the energy per baryon for a given density $\rho$ 
(see Refs.\ \cite{sprung72:a,day78:a,pandharipande79:a,wiringa88:a} for 
details):
\beqn
      E_v(\rho) \; = \; {\rm min} \;\;\; \left\{
{ {<\, \Psi_v\, |\, H\, |\, \Psi_v\, >}
               \over{<\, \Psi_v\, |\, \Psi_v\, >} } \right\}
\; \; \geq \; \; E_0 \; .
\label{eq:evar}
\eeqn
As indicated, $E_v$ constitutes an upper bound to the ground-state energy
$E_0$. The energy density $\epsilon(\rho)$ and pressure $P(\rho)$
are obtained from Eq.\ (\ref{eq:evar}) by 
\beqn
         \epsilon(\rho) \; = \; \rho \; [E_v(\rho)\, + \, m] \; ,
\qquad \quad
            P(\rho)\; = \; \rho^2 \; { {\partial\;\;\, }\over
         {\partial \rho} }\; E_v(\rho)\;  ,
\label{eq:f52}
\eeqn
which leads to the \eos in the form $P(\epsilon)$ used for the
star structure calculations here.

\subsubsection{Relativistic approach}\label{sssec:relativ}

As discussed elsewhere \cite{weber91:d,glen85:b}, the lagrangian governing
the dynamics of many-baryon (-lepton) neutron-star matter has the
following structure:
\begin{eqnarray}
{\cal L}(x) \, &=&  \, \sum_{B=p,n,\Sigma^{\pm,0},\Lambda,\Xi^{0,-},
        \Delta^{++,+,0,-} }  {\cal L}^0_B(x) \label{eq:f31} \\
  &+& \sum_{M=\sigma,\omega,\pi,\varrho,\eta,\delta,\phi}
     \left\{ {\cal L}^0_M(x) \;+\; \sum_{B=p,n,...,\Delta^{++,+,0,-}}
  {\cal L}^{\rm Int}_{B,M} (x) \right\} 
  + \sum_{\lambda=e^-,\mu^-} \;{\cal L}_\lambda (x)\;. \nonumber
\end{eqnarray} The subscript $B$ runs over all baryon species that become
populated in dense star matter, which are listed in Table \ref{tab:bary}. The
nuclear forces are mediated by that collection of scalar, vector, and isovector
mesons ($M$) that is used for the construction of relativistic
one-boson-exchange potentials \cite{holinde72:a,machleidt87:a}.
\begin{table}
\caption[Masses, electric charges and quantum numbers of the baryons
included in the determination of the \eos of neutron star matter]{Masses,
electric charges ($q_B$) and quantum numbers
(spin ($J_B$), isospin ($I_B$), strangeness ($S_B$), hypercharge
($Y_B$), third component of isospin ($I_{3B}$)) of the baryons
included in the determination of the \eos of neutron star matter.}
\label{tab:bary}
\begin{center}
\begin{tabular}{lrcrrrrr}\hline
Baryon ($B$) &$m_B\;[$MeV$]$ &$J_B$ &$I_B$ &$S_B$ &$Y_B$ &$I_{3B}$ &$q_B$\\
                                                        \hline \hline
$n$ &939.6    &$1/2$ &$1/2$ &0     &1    &$-1/2$ &0            \\
$p$ &938.3    &$1/2$ &$1/2$ &0     &1    &$1/2 $ &1            \\
$\Sigma^+$    &1189  &$1/2$ &1     &$-1$ &0      &1      &1    \\
$\Sigma^0$    &1193  &$1/2$ &1     &$-1$ &0      &0      &0    \\
$\Sigma^-$    &1197  &$1/2$ &1     &$-1$ &0      &$-1$   &$-1$ \\
$\Lambda $    &1116  &$1/2$ &0     &$-1$ &0      &0      &0    \\
$\Xi^0$       &1315  &$1/2$ &$1/2$ &$-2$ &$-1$   &$1/2$  &0    \\
$\Xi^-$       &1321  &$1/2$ &$1/2$ &$-2$ &$-1$   &$-1/2$ &$-1$ \\
$\Delta^{++}$ &1232  &$3/2$ &$3/2$ &0    &1      &$3/2$  &2    \\
$\Delta^{+}$  &1232  &$3/2$ &$3/2$ &0    &1      &$1/2$  &1    \\
$\Delta^{0}$  &1232  &$3/2$ &$3/2$ &0    &1      &$-1/2$ &0    \\
$\Delta^{-}$  &1232  &$3/2$ &$3/2$ &0    &1      &$-3/2$ &$-1$ \\  \hline
\end{tabular}
\end{center}
\end{table} The equations of motion resulting from Eq.\ (\ref{eq:f31}) for the
various baryon and meson field operators (given in \cite{weber91:d,glen85:b})
are to be solved subject to the conditions of electric charge neutrality,
$\rho^{\rm el}_{\rm tot}=\rho^{\rm el}_{\rm Bary} +\rho^{\rm el}_{\rm
  Lep}\equiv0$, which leads to
\begin{eqnarray}
  \sum_B q_B \; (2J_B+1)\; {{p_{F,B}^3}\over{6\pi^2}}\; - \; \biggl[
  \sum_{\lambda=e,\mu} { {p_{F,\lambda}^3}\over{3\pi^2}} \, + \,
  \varrho_\pi\;\Theta(\mu^\pi-m_\pi) \biggr]\; = \; 0 \; ,
\label{eq:chargeneut}
\end{eqnarray} and $\beta$ (chemical) equilibrium \cite{glen85:a,weber89:e}, 
\begin{equation}
\mu^B = \mu^n - q_B \, \mu^e \; ,
\label{eq:mub}
\end{equation} where $\mu^n$ and $\mu^e$ denote the chemical potentials of
neutrons and electrons, whose knowledge is sufficient to find the chemical
potential of any other baryon of electric charge $q_B$ present in the system.
To solve the equations of chemically equilibrated many baryon matter we apply
the Greens function method, which is outlined next.

The starting point is the Martin-Schwinger hierarchy of coupled Greens
functions \cite{martin59:a,wilets79:a}.  In the lowest order, the
Martin-Schwinger hierarchy can be truncated by factorizing the four-point
Greens function $g_2(1,2;1'2')$ [unprimed (primed) arguments refer to ingoing
(outgoing) particles] into a product of two-point Greens functions $g$ $[\equiv
g_1(1';1')]$.  This leads to the well-known relativistic Hartree (i.e.,
mean-field) and Hartree-Fock approximations.  The T-matrix approximation (also
known as $\Lambda$ or ladder approximation), which goes beyond this, truncates
the Martin-Schwinger hierarchy by factorizing the six-point Greens function
$g_3(123;1'2'3')$ into products of four- and two-point functions by which
dynamical two-particle correlations in matter - which are connected with the
two-body potential (denoted $v$) - are taken into account. The main problem
which one encounters hereby is the calculation of the effective scattering
matrix (effective two-particle potential) in matter, $T$, which satisfies 
\beqn
T\; =\; v - v^{\rm ex} + \int v \; \Lambda \; T \; .
\label{eq:tmatrix}
\eeqn We have restricted ourselves to the so-called $\Lambda^{00}$
approximation, for which the nucleon-nucleon propagator is given by
the product of two free two-point Greens functions, i.e.,
$\Lambda=\Lambda^{00}\equiv i g^0 g^0$, and the relativistic
Brueckner-Hartree-Fock (RBHF) approach, where both intermediate
baryons feel the nuclear background medium and scatter only into
states outside the Fermi sea \cite{huber93:a,huber94:a,huber95:a}.
The basic input quantity in Eq.\ (\ref{eq:tmatrix}) is the
nucleon-nucleon interaction in free space as derived, for example, in
the Bonn meson-exchange model \cite{machleidt87:a}. We have adopted
the latest version of this interaction together with Brockmann's
potentials A--C to compute the T matrix in neutron matter up to
several times nuclear matter density
\cite{weber90:a,huber93:a,huber94:a,huber95:a}.  The important feature
of such meson-exchange models is that the potential parameters are
adjusted to the two-body nucleon-nucleon scattering data and the
properties of the deuteron, whereby (in this sense) a parameter-free
treatment of the many-body problem is achieved.  The ``Born''
approximation of $T$ sums the various meson potentials of the
nucleon-nucleon interaction in free space, i.e., \beqn <1 2\mid v \mid
1' 2'>\; = \; \sum_{M=\sigma,\omega,\pi,\rho,\eta,\delta, \phi} \;
\delta^4_{11'} \;\; \Gamma^M_{11'} \;\; \Delta^M_{12} \;\;
\Gamma^M_{22'} \;\; \delta^4_{22'} \; ,
\label{eq:vborn}
\eeqn and thus neglects dynamical nucleon-nucleon correlations.  It is
this approximation which leads to the relativistic Hartree and
Hartree-Fock approximations
\cite{weber91:d,poschenrieder88:a,weber88,weber89:e}.  The symbol
$\Gamma^M$ in Eq.\ (\ref{eq:vborn}) stands for the various
meson-nucleon vertices, and $\Delta^M$ denotes the free meson
propagator of a meson of type $M$.

The nucleon self-energy (effective one-particle potential) is obtained 
from the T matrix as \cite{weber91:d,weber89:a}
\beqn  
\Sigma^B \; = \; i \; \sum_{B'} \int \; \left[ \; {\rm tr}\; 
\left( T^{BB'} \;g^{B'} \right) - T^{BB'}\;g^{B'}\; \right] \; ,
\label{eq:sigmal}
\eeqn
where $B'=p,n,\Sigma^{\pm,0},\Lambda,\Xi^{0,-}, \Delta^{++,+,0,-}$
sums all the charged baryon states whose threshold 
densities are reached in the cores of neutron star models.
\begin{figure}[tb]
\begin{center}
\leavevmode
\psfig{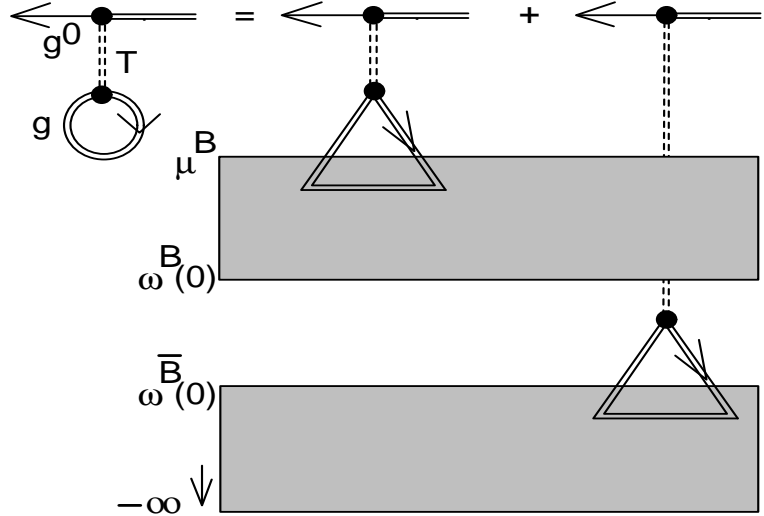}
\caption[Graphical representation of the self-energy contributions arising
from both the Fermi and Dirac seas]{Graphical representation of the 
self-energy contributions (second term in Eq.\ (\ref{eq:dyson}))
arising from both the Fermi sea, i.e.,\ nuclear 
matter consisting of filled baryon states of energies 
$\omega^B(0) \leq \omega^B(p) \leq \mu^B$ (upper graph),
as well as Dirac sea (states
$-\infty < \omega^{\bar B}(p) \leq \omega^{\bar B}(0)$, lower graph). 
The former (latter) lead to so-called medium (vacuum) polarization 
contributions \cite{serot86:a,horowitz87:a}.\label{fig:fermisea}}
\end{center}
\end{figure}
The baryon propagators in Eq.\ (\ref{eq:sigmal}) 
 are given as the solutions of Dyson's equation,
\beqn
      g^B\; = \; g^{0\,B} \; + \; g^{0\,B} \; \Sigma^B(\{ g^{B'} \})
      \; g^B \; ,
\label{eq:dyson} 
\eeqn which terminates the set of equations that are to be solved
self-consistently.  The diagrammatic representation of the second term
in Dyson's equation, $g^{0\,B} \Sigma^B g^B$, is given in Fig.\ 
\ref{fig:fermisea}. There, single lines denote the free propagator,
$g^{0\,B}$, and double lines refer to the self-consistent propagator
in matter, $g^B$. This term corrects the free propagators (first term)
for medium effects arising from the Fermi sea of filled baryon states,
i.e., the nuclear-matter medium (upper shaded area in Fig.\ 
\ref{fig:fermisea}), within which the baryons
move.\footnote{Contributions coming from the lower shaded area account
  for vacuum polarization corrections. The so-called no-sea
  approximation, which has been applied for the determination of most
  of the \eoss presented in Sect.\ \ref{ssec:models}, neglects such
  corrections.  A critical discussion of the influence of vacuum
  renormalization on the \eos of high-density matter has been
  performed in Ref.\ \cite{glen89:a493}. It was found that these have
  negligible influence on the \eos up to densities of at least ten
  times normal nuclear matter density, provided the coupling constants
  are tightly constrained by the saturation properties of nuclear
  matter. In the present paper, vacuum polarization contributions are
  contained in \eoss denoted $\egth$ and $\egthpi$ (see Table
  \ref{tab:eos}).} Obviously these vanish for baryons propagating in
free space, for which $\Sigma^B \equiv 0$.  Note that the
determination of $g^B$ from Dyson's equation leads to a
self-consistent treatment of the coupled matter equations
(\ref{eq:tmatrix})--(\ref{eq:dyson}).

The \eos follows from the stress-energy density tensor, ${\cal
  T}_{\mu\nu}$, of the system as 
\beqn E(\rho) \; &=& \; <\,{\cal
  T}_{00}\,> / \rho \; - \; m \; , \qquad\qquad {\rm where} \\ 
\label{eq:engpn}
{\cal T}_{\mu\nu}(x)\; &=&\; \sum_{\chi=B,\lambda} 
\partial_\nu\Psi_\chi(x) \; {{\partial
{\cal L}(x)}  \over{\partial \;(\partial^\mu\,\Psi_\chi(x))}}
\;-\;g_{\mu\nu} \; {\cal L}(x)\; .
\label{eq:emtensor}
\eeqn The sum in the latter equation sums the contributions coming
from the baryons and leptons ($\lambda=e,\mu$). The quantity ${\cal
  L}$ denotes the lagrangian given in Eq.\ (\ref{eq:f31}) (see Ref.
\cite{weber89:e} for details). The pressure is obtained from $E(\rho)$
via Eq.\ (\ref{eq:f52}).

\subsection{Models for the nuclear \eos}\label{ssec:models}

A representative collection of nuclear \eoss that are determined in the 
framework of non-relativistic Schroedinger theory and relativistic nuclear
field theory is listed in Table \ref{tab:eos}. A few of them are graphically 
shown in Fig.\ \ref{fig:eos}, where the pressure is plotted as a 
function of energy density (in units of the density of normal nuclear matter, 
\begin{table}[tb]
\caption[Representative collection of nuclear \eoss]{Nuclear \eoss applied
for the construction of models of general relativistic (rotating) neutron star 
models.}
\label{tab:eos}
\begin{center}
\begin{tabular}{cclll} \hline
   Label  &      &EOS   &Description (see text) &Reference   \\  \hline \hline
  \multicolumn{5}{c}{Relativistic field theoretical equations of state} \\
  \cline{1-5}
1 &  &G$_{300}$                   &H,\,$K$=300   &\cite{glen89:a493} \\
2 &  &HV                          &H,\,$K$=285   &\cite{glen85:b,weber89:e} \\
3 &  &G$^{\rm DCM2}_{{\rm B}180}$  &Q,\,$K$=265,\,$B^{1/4}=180$
                                         &\cite{glen91:c,glen91:pt} \\
4 &  &G$^{\rm DCM2}_{265}$        &H,\,$K$=265  &\cite{glen91:b} \\
5 &  &G$^\pi_{300}$               &H,\,$\pi$,\,$K$=300 &\cite{glen89:a493} \\
6 &  &G$^\pi_{200}$               &H,\,$\pi$,\,$K$=200 &\cite{glen86:b} \\
7 &  &$\Lambda^{00}_{\rm Bonn}+{\rm HV}$ &H,\,$K$=186  &\cite{weber90:a} \\
8 &  &G$^{\rm DCM1}_{225}$         &H,\,$K$=225  &\cite{glen91:b} \\
9 &  &G$^{\rm DCM1}_{{\rm B}180}$  &Q,\,$K$=225,\, $B^{1/4}=180$
                                        &\cite{glen91:c,glen91:pt}  \\
10&  &HFV                         &H,\,$\Delta$,\,$K$=376 &\cite{weber89:e} \\
11&  &$\Lambda^{\rm RBHF}_{\rm Bro}+{\rm HFV}$
                                 &H,\,$\Delta$,\,$K$=264
             &\cite{huber93:a,huber94:a,huber95:a} \\
\cline{1-5}
  \multicolumn{5}{c}{Non-relativistic potential model equations of state} \\
\cline{1-5}
12&  &{\small BJ(I)}                    &H,\,$\Delta$ &\cite{bethe74:a} \\
13&  &{\small WFF(UV$_{14}$+TNI)}       &NP,\,$K$=261 &\cite{wiringa88:a}\\
14&  &FP(V$_{14}$+TNI)                  &N,\,$K$=240  &\cite{friedman81:a}\\
15&  &{\small WFF(UV$_{14}$+UVII)}      &NP,\,$K$=202 &\cite{wiringa88:a}\\
16&  &{\small WFF(AV$_{14}$+UVII)}      &NP,\,$K$=209 &\cite{wiringa88:a}\\
17&  &{\small MS94}                     &N,\,$K$=234 
          &\cite{myers90:a,myers91:a,myers94:a,myers94:c}  \\
                                                                    \hline
\end{tabular}
\end{center}
\end{table}
$\epsilon_0=140~\mevt$). This collection of \eoss has been applied for the 
construction of models of general relativistic (rotating) compact star models, 
which will be presented in Sect.\ \ref{sec:properties}. 
The specific properties of these \eoss are described in 
Table \ref{tab:eos}, where the following abbreviations are used:
N = pure neutron;
NP = $n,\,p$, leptons;
$\pi$ = pion condensation;
H = composed of $n,\,p$, hyperons ($\Sigma^{\pm,0},\,\Lambda,\,\Xi^{0,-}$), and
leptons;
$\Delta$ = $\Delta_{1232}$-resonance;
Q = quark hybrid composition, i.e., $n,\,p$, hyperons in equilibrium with 
$u,\,d,\,s$-quarks, leptons;
$K$ = incompressibility (in MeV);
$B^{1/4}$ = bag constant (in MeV).
Not all \eoss of our collection account for neutron matter in 
full $\beta$ equilibrium (i.e., entries 13--17). 
These models treat neutron star matter
as being composed of only neutrons, or neutrons and protons in
equilibrium with leptons, which is however
not the ground-state of neutron star matter predicted by theory
\cite{glen85:b,bethe74:a,pandharipande71:a}.
As an example of such an \eosp, we exhibit the $\efp$ model in 
Fig.\ \ref{fig:eos}. The relativistic \eoss account for all baryon states that 
become populated in dense star models constructed from them. 
As representative examples for the relativistic \eossp, we show the
HV, HFV, $\egth$, and $\egdcma$ models in Fig.\ \ref{fig:eos}.
A special feature of the latter \eos is that it also 
(as $\egdcmc$, which is not shown in Fig.\ \ref{fig:eos})
accounts for 
the possible transition of baryon matter to quark matter.
One clearly sees in Fig.\ \ref{fig:eos} the softening of the \eosp,
i.e., reduction of pressure for a given density, at  
$\eps \gsim (2-3)\,\epsnm$ which is caused by the onset of baryon population 
and/or the transition of baryon matter to quark matter.
The stiffer behavior of HFV in comparison with HV at high densities has its 
origin in the exchange (Fock) contribution that is contained in the former
\eosp.
An inherent feature of the relativistic \eoss is that they
do not violate causality, i.e., the velocity of sound, given by
$v_s=c\,\sqrt{dP/d\eps}$, is smaller than the velocity of light ($c$) 
at all densities, which is not the case for the 
non-relativistic models for the \eos (cf.\ Table \ref{tab:bulk17}).
Among the latter only the $\ewut$ \eos does not violate
causality up to densities relevant for the construction of models of
neutron stars from it.

The nuclear matter properties at saturation density 
related to our collection of 
\eoss are summarized in Table \ref{tab:bulk17}.
The listed quantities are: binding energy of normal nuclear matter at
saturation density, $E/A$; compression modulus, $K$;
effective nucleon mass, $M^*$ $(\equiv m^*/m$, where
$m$ denotes the nucleon mass); asymmetry energy, $a_{\rm sy}$.
With the exception of $\ebnhv$ and $\Lambda^{\rm RBHF}_{\rm Bro}+{\rm HFV}$, 
the coupling constants of the relativistic \eoss are
determined such that these saturate infinite nuclear matter at
densities in the range 0.15 to 0.16~$\fmmt$ for a binding energy per nucleon
of about $- 16$ MeV. For the $\ebnhv$ and 
$\Lambda^{\rm RBHF}_{\rm Bro}+{\rm HFV}$ \eoss the saturation 
properties are determined
by respectively the relativistic Bonn and Brockmann meson-exchange models for
the nucleon-nucleon interaction whose parameters are determined by the free
\begin{table}[tb]
\caption[Survey of nuclear matter properties]{Nuclear matter properties of
the \eoss used in this work.\label{tab:bulk17}}
\begin{center}
\begin{tabular}{clllllll} \hline
{\small Label} &{\small EOS} &$E/A$  &$\rho_0$  &$K$  &$M^*$
&$a_{\rm sy}$  &$\epsilon/\epsilon_0\,^\dagger$  \\
     &
     &$[\rm MeV]$  &[fm$^{-3}]$  &$[\rm MeV]$  &$[\rm MeV]$
     &$[\rm MeV]$ &  \\ \hline \hline
1   &G$_{300}$                &$-16.3$   &0.153  &300    &0.78   &32.5 &$-$ \\
2   &HV                       &$-15.98$  &0.145  &285    &0.77   &36.8 &$-$ \\
3   &$\egdcmc$                &$-16.0$   &0.16   &265    &0.796  &32.5 &$-$ \\
4   &$\egdcmd$                &$-16.0$   &0.16   &265    &0.796  &32.5 &$-$ \\
5   &G$^\pi_{300}$            &$-16.3$   &0.153  &300    &0.78   &32.5 &$-$ \\
6   &G$^\pi_{200}$            &$-15.95$  &0.145  &200    &0.8    &36.8 &$-$ \\
7   &$\ebnhv$                  &$-11.9$  &0.134  &186    &0.79   &     &$-$ \\
8   &$\egdcmb$                 &$-16.0$  &0.16   &225    &0.796  &32.5 &$-$ \\
9   &$\egdcma$                 &$-16.0$  &0.16   &225    &0.796  &32.5 &$-$ \\
10  &HFV                       &$-15.54$ &0.159  &376    &0.62   &30   &$-$ \\
11  &$\Lambda^{\rm RBHF}_{\rm Bro}+{\rm HFV}$   &$-14.81$ &0.170 &264 &0.66
   &32     &$-$ \\
12  &{\small BJ(I)}              &         &       &       &       &    &23.1\\
13  &{\small WFF(UV$_{14}$+TNI)} &$-16.6$    &0.157  &261   &0.65 &30.8 &14 \\
14  &{\small FP(V$_{14}$+TNI)}   &$-16.00$   &0.159  &240   &0.64 &     &5.6 \\
15  &{\small WFF(UV$_{14}$+UVII)}&$-11.5$    &0.175  &202   &0.79 &29.3 &6.5 \\
16  &{\small WFF(AV$_{14}$+UVII)}&$-12.4$    &0.194  &209   &0.66 &27.6 &7.2 \\
17  &{\small MS94}               &$-16.04$   &0.161  &234   &     &32.0 &13  \\
\hline
\makebox[0.05in][l]{$^\dagger$ \parbox[t]{4.9in}{Energy density in units of
normal nuclear matter density beyond which the
velocity of sound in neutron matter becomes larger (superluminal) than
the velocity of light. The symbol ``$-$'' indicates that causality is
not violated.}}
 & & & & & & & \\
\end{tabular}
\end{center}
\end{table} nucleon-nucleon scattering problem and the properties of
the deuteron (parameter-free treatment). The influence of dynamical
two-particle correlations calculated from the scattering matrix leads
for these two \eoss to a relatively soft behavior in the vicinity of
the saturation density. This is indicated by the rather small
compression moduli $K$ related to these \eossp. All non-relativistic
\eoss of our collection, which are determined in the framework of the
variational method outlined in Sect.\ \ref{sssec:nonr}, contain the
impact of dynamical two-particle correlations in matter, too.  The
correlations are calculated for different hamiltonians. With the
exception of BJ(I) and MS94, the calculations are performed for the
Urbana and Argonne two-nucleon potentials $V_{14}$, ${\rm UV}_{14}$
\cite{schiavilla86:a} and ${\rm AV}_{14}$ \cite{wiringa84:a},
respectively, supplemented by different models for the three-nucleon
interaction.  These are the density-dependent three-nucleon
interaction of Lagaris and Pandharipande, TNI \cite{lagaris81:a}, and
the Urbana three-nucleon model, UVII \cite{schiavilla86:a}.  One sees
that nuclear matter is underbound by $\approx 4$ MeV for two of these
\eossp. The corresponding saturation densities are in the range of
0.17 to 0.19~$\fmmt$, thus nuclear matter saturates at somewhat too
large densities for these \eossp.  (The empirical saturation density
is $\approx 0.15\;\fmmt$ \cite{myers69:a}.) \Eoss labeled 13 and 14
lead to binding energies and saturation densities that are in good
agreement with the empirical values which has its origin in the
density-dependent three-nucleon interaction TNI.\footnote{It is well
  known that two-particle correlations alone fail in reproducing the
  empirical values of binding energy and saturation density. In this
  case the saturation points calculated from the standard
  Brueckner-Hartree-Fock
  \protect{\cite{sprung72:a,banj71:a,mahaux81:a}} and non-relativistic
  T matrix approximations \protect{\cite{fiset72:a,kim74:a,weber85}}
  for different nucleon-nucleon interactions fall in a narrow band,
  often called the Coester band. It appears likely that this band
  would contain the calculated saturation point for any realistic
  nucleon-nucleon interaction \protect{\cite{coester70:a,wong72:a}}.}
Interesting is the shift of the saturation density obtained for
$\ebnhv$ and $\Lambda^{\rm RBHF}_{\rm Bro}+{\rm HFV}$ -- relative to
non-relativistic treatments that account for dynamical correlations
too -- toward smaller densities which is caused by relativity
\cite{poschenrieder88:a,serot86:a,celenza81:a,celenza86:a,malfliet87:a}.

The four \eoss $\egdcmb$, $\egdcmd$, $\egdcma$, and $\egdcmc$, which  are
based on the relativistic lagrangian of Zimanyi and Moszkowski, have only 
recently been determined \cite{glen91:pt,weber91:a}.
The transition of confined hadronic matter into quark matter 
is taken into account in \eoss $\egdcma$ (Fig.\ \ref{fig:eos})
and $\egdcmc$. Here 
a bag constant of $B^{1/4}=180$ MeV has been used for the determination 
\begin{figure}[tb]
\begin{center}
\leavevmode
\psfig{figure=eoss.bb,width=7.0cm,height=8.0cm,angle=90}
\caption[Graphical illustration of the \eoss HV, HFV, $\efp$, $\egth$, and
$\egdcma$]{Graphical illustration of the \eoss HV, HFV, $\efp$, $\egth$, and
$\egdcma$.\label{fig:eos}}
\end{center}
\end{figure} of the transition of baryon matter into quark matter, which places
the energy per baryon of strange matter at 1100 MeV, well above the energy per
nucleon in $^{56}{\rm Fe}$ ($\approx$ 930 MeV).  Most interestingly, the
transition to quark matter sets in already at a density $\eps = 2.3\,\epsnm$
\cite{glen91:c,glen91:pt}, which lowers the pressure relative to confined
hadronic matter.  The mixed phase of baryons and quarks ends, i.e. the pure
quark phase begins, at $\eps \approx 15 \, \epsnm$, which is larger than the
central density encountered in the maximum-mass star model constructed from
this \eosp.  We stress that these density thresholds are rather different from
those computed by other authors in earlier investigations.  The reason for this
lies in the realization that the transition between confined hadronic matter
and quark matter takes place subject to the conservation of baryon and electric
charge. Correspondingly, there are two chemical potentials, and the transition
of baryon matter to quark matter is to be determined in three-space spanned by
pressure and the chemical potentials of the electrons and neutrons.  The only
existing investigation which accounts for this properly has been performed by
Glendenning \cite{glen91:c,glen91:pt}.  Further important differences between
the determination of $\egdcma$ and earlier (and thus inconsistent) treatments
concern the description of the dense interior of compact stars\footnote{If the
  dense core may be converted to quark matter
  \cite{glen89:j,glen89:k,ellis91:a}, it must be strange quark matter, since
  3-flavor quark matter has a lower energy per baryon than 2-flavor. And just
  as is the case for the hyperon content of neutron stars, strangeness is not
  conserved on macroscopic time scales. Many of the earlier discussions
  \cite{glen89:j,glen89:k,ellis91:a,baym76:a,keister76:a,chap77:a,fech78:a,bethe87:a,serot87:a}
  have treated the neutron star as pure in neutrons, and the quark phase as
  consisting of the equivalent number of $u$ and $d$ quarks. However neither is
  pure neutron matter the ground state of a star nor is a mixture of $n_d = 2\,
  n_u$! In fact it is a highly excited state, and will quickly weak decay to an
  approximate equal mixture of $u$, $d$, $s$ quarks.} and the approximation of
the mixed phase as two components which are separately charge neutral.

\section{Observed neutron star properties}\label{sec:obsd}

The global neutron star properties such as masses, rotational frequencies,
radii, moments of inertia, redshifts, etc.\ are known to be sensitive to the
adopted microscopic model for the nucleon-nucleon interaction or, in other
words, to the nuclear \eos \cite{arnett77:a}. Thus, by means of comparing the
theoretically determined values for these quantities with observed ones one may
hope to learn about the physical behavior of matter at super-nuclear densities.
In the following we briefly summarize important star properties.

\subsection{Masses}\label{ssec:masses}

The gravitational mass is of special importance since it can be inferred
directly from observations of X-ray binaries and binary pulsars (e.g., the
Hulse-Taylor radio pulsar PSR 1913+16 \cite{taylor89:a}).  Rappaport and Joss
were the first who deduced neutron star masses
\begin{figure}[tb]
\begin{center}
\leavevmode
\psfig{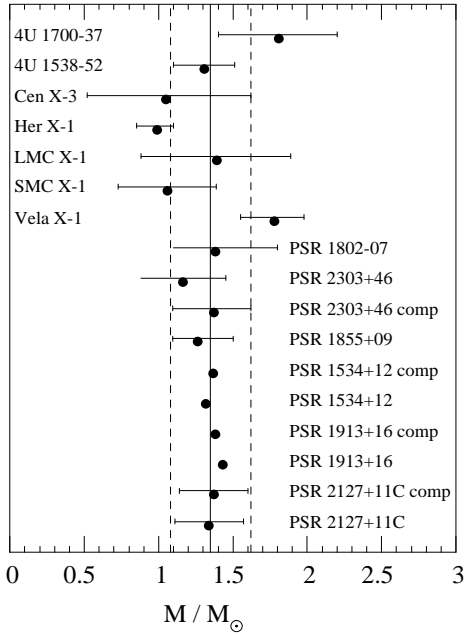}
\caption[Distribution of observed neutron star masses]{Distribution
of observed neutron star masses \cite{thorsett93:a}.}\label{fig:nsmasses}
\end{center}
\end{figure}
for six X-ray binaries \cite{rappaport83:a}. A reexamination of these 
masses became possible owing to the improved determinations  of 
orbital parameters \cite{nagase89:a}. Improved values are
shown in Fig.\ \ref{fig:nsmasses}.
The objects 3U1700--37 and Cyg~X--1 are 
two non-pulsating X-ray
binaries. It is expected that at least one of these objects,
likely Cyg X-1, is a black hole. Very remarkable is the 
extremely accurately determined mass of the Hulse-Taylor
binary pulsar, PSR 1913+16, which is given by 
$1.444\pm 0.003$ \cite{taylor89:a}. 

In summary, Fig.\ \ref{fig:nsmasses} suggests that  
the most probable value of neutron star masses, as derived  
from observations of
binary X-ray pulsars,  is close to about $1.4\,\msun$, and that  
 the masses of individual neutron stars 
are likely to be in the range $1.1 \lsim M/\msun \lsim 1.8$ 
\cite{nagase89:a}.

\subsection{Rotational frequencies of fast pulsars}\label{ssec:freq}

The rotational periods of fast pulsars provide conditions on the \eos when 
combined with the mass constraint \cite{glen90:ss}. As already 
mentioned in Sect.\ \ref{sec:intro}, the fastest so far observed
pulsars have rotational periods of 1.6 ms (i.e., 620 Hz).
The successful model for the nuclear \eosp, therefore, must account for 
rotational neutron star periods of at least $P = 1.6$ ms as well as 
masses that lie in range listed in Sect.\ \ref{ssec:masses}.

\subsection{Radii}\label{ssec:radii}

Direct radius determinations for neutron stars do not exist. However,
combinations of data of 10 well-observed X-ray bursters with special
theoretical assumptions lead Van Paradijs \cite{paradijs78:a} to the
conclusion that the emitting surface has a radius of about 8.5 km. This
value, as pointed out in \cite{shapiro83:a}, may be underestimated by a
factor of two. Fujimoto and Taam \cite{fujimoto86:a} derived from the
observational data of the X-ray burst source MXB 1636--536, under rather 
uncertain theoretical assumptions, a neutron star mass and radius of 
$1.45\,\msun$ and 10.3 km. An error analysis lead them to predicting
mass and radius ranges of 1.28 to $1.65\,\msun$ and 9.1 to 11.3 km, 
respectively. When comparing these values with computed neutron star data,
however, one should be aware of the fact that burster are suspected, 
but not known, to be neutron stars. 

\subsection{Moment of inertia}\label{ssec:moi}

Another global neutron star property is the \moip, $I$.  Early
estimates of the energy-loss rate from pulsars \cite{ruderman72:a}
spanned a wide range of $I$, i.e., $7\times 10^{43} < I < 7\times
10^{44}~\gcmt$.  From the luminosity of the Crab nebula ($\sim 2-4
\times 10^{38}~ {\rm erg/sec}$), several authors have found a lower
bound on the moment of inertia of the pulsar given by $I \gsim
4-8\times 10^{44}~ \gcmsq$ \cite{baym75:a,trimble70:a,borner73:a}.

\subsection{Redshift}\label{ssec:redshift}

Finally we mention the neutron star redshift, $z$.
Liang \cite{liang86:a} has considered the neutron star redshift data base
provided by measurements of $\gamma$-ray burst redshifted annihilation
lines in the range $300-511$ keV. These bursts have widely been interpreted
as gravitationally redshifted 511 keV $e^\pm$ pair annihilation lines from the
surfaces of neutron stars. From this he showed that there is tentative
evidence (if the interpretation is correct) to support a neutron star 
redshift range of $0.2 \leq z \leq 0.5$,
with the highest concentration in the narrower range $0.25 \leq z \leq 0.35$.
A particular role plays the source of the 1979 March 5 $\gamma$-ray burst 
source, which has been identified with SNR N49 by its position. From the 
interpretation of its emission, which has a peak at $\sim 430$ keV,
as the 511 keV $e^\pm$ annihilation line \cite{brecher80:a,ramaty81:a}
the resulting gravitational redshift has a value of $z=0.23 \pm 0.05$.

\section{Properties of neutron star models}\label{sec:properties}

\subsection{Non-rotating star models}\label{sec:nonrotst}

The structure of spherical neutron stars is determined by the
Oppenheimer-Volkoff equations \cite{oppenheimer39,misner73:a},
\begin{eqnarray}
{{dP}\over{dr}} = - \, \frac{\epsilon(r)\, m(r) \, 
\left[1 + P(r)/\epsilon(r) \right]
\,  \left[ 1 + 4 \pi r^3 P(r)/m(r) \right]}
{r^2 \, \left[1 - 2 m(r)/r \right]} \; ,
\label{eq:f28}
\end{eqnarray}
which describe a compact stellar configuration in hydrostatic equilibrium.
(Here we use units for which the gravitational constant and 
velocity of light are $G=c=1$. Hence $\msun = 1.5$ km.)
The boundary condition reads $P(r=0) \equiv P_c = P(\epsilon_c)$, where
$\eps_c$ denotes the energy density at the star's center, which constitutes
the free parameter that is to be specified when solving the above 
differential equation for a given \eosp. The latter determines 
$P_c$ and the energy density for all pressure values $P<P_c$.
The pressure is to be computed
out to that radial distance where $P(r)=0$, which determines the star's
radius $R$, i.e., $P(r=R)=0$.
The mass contained in a sphere of radius $r~(\leq R)$, denoted by $m(r)$,
follows from $\epsilon(r)$ as 
\begin{eqnarray}
m(r) = 4 \pi \int\limits^{r}_0 dr'\; r'^2 \; \epsilon(r') \; .
\label{eq:f29}
\end{eqnarray}
The star's total (gravitational) mass is given by $M\equiv m(R)$. 

Figure \ref{fig:ovme1} exhibits the gravitational mass of non-rotating
neutron stars as a function of central energy density for a sample of
\eoss of Table \ref{tab:eos}.  Each star sequence is shown up to
densities that are slightly larger than those of the maximum-mass star
(indicated by tick marks) of each sequence. Stars beyond the mass peak
are unstable against radial oscillations and thus cannot exist stably
(collapse to black holes) in nature.
\begin{figure}[tb]
\begin{center}
\parbox[t]{6.5 cm}
{\leavevmode
\psfig{figure=msphec.bb,width=6.0cm,height=7.0cm,angle=90}
{\caption[Non-rotating gravitational mass versus central
density]{Non-rotating neutron star mass as a function of central 
density.}\label{fig:ovme1}}}
\ \hskip1.4cm   \
\parbox[t]{6.5cm}
{\leavevmode
\psfig{figure=msphz.bb,width=6.0cm,height=7.0cm,angle=90}
{\caption[Non-rotating star mass versus gravitational redshift]{Non-rotating
neutron star mass as a function of redshift.}\label{fig:ovmz1}}}
\end{center}
\end{figure} One sees that all \eoss are able to support non-rotating
neutron star models of gravitational masses $M \geq M({\rm PSR}~1913+16)$.
On the other hand, rather massive stars of say $M\gsim 2\,M_\odot$ can
only be obtained for those models of the \eos that exhibit a rather
stiff behavior at super nuclear densities (cf.\ Fig.\ \ref{fig:eos}).
The largest maximum-mass value, $M=2.2\,\msun$, is obtained for HFV,
which is caused by the exchange term contained in that model.
Knowledge of the maximum-mass value is of great importance for two
reasons. Firstly, quite a few neutron star masses are known (Sect.\ 
\ref{sec:obsd}), and the largest of these imposes a lower bound on the
maximum-mass of a theoretical model. The current lower bound is about
$1.56\,\msun$ [neutron star 4U\,0900--40 ($\equiv$ Vela X--1)], which,
as we have just seen, does not set a too stringent constraint on the
nuclear \eosp. The situation could easily change if an accurate future
determination of the mass of neutron star 4U\,0900--40 (see Fig.\ 
\ref{fig:ovme1}) should result in a value that is close to its present
upper bound of $1.98\,\msun$.  In this case most of the \eoss of our
collection would be ruled out.  The second reason is that the maximum
mass can be useful in identifying black hole candidates
\cite{ruffini78:a,brown94:a,bethe95:a}.  For example, if the mass of a
compact companion of an optical star is determined to exceed the
maximum mass of a neutron star it must be a black hole. Since the
maximum mass of stable neutron stars in our theory is $ 2.2\,\msun$,
compact companions being more massive than that value are predicted to
be black holes.

The neutron star mass as a function of gravitational redshift, defined as
\begin{eqnarray}
    z = \frac{1}{\sqrt{1 - 2 M / R}} \, - \, 1 \; ,
\label{eq:f274}
\end{eqnarray}
is shown in Fig.\ \ref{fig:ovmz1} for the same sample of \eoss
as in Fig.\ \ref{fig:ovme1}. One sees that the maximum-mass
stars have redshifts in the range $0.4\lsim z \lsim 0.8$,
depending on the softness (stiffness) of the \eosp.
Neutron stars of typically $M\approx 1.5 \,\msun$ (e.g., PSR 1913+16) 
are predicted
to have redshifts in the considerably narrower range $0.2 \leq z \leq 0.32$.
The solid rectangle covers
masses and redshifts in the ranges of $1.30 \leq M/\msun
\leq 1.65$ and $0.25 \leq z \leq 0.35$, respectively. 
As outlined in Sects.\ \ref{ssec:radii} and \ref{ssec:redshift},
the former range has been determined
from observational data of X-ray burst source MXB~1636--536
\cite{fujimoto86:a}, while the latter is based on the neutron star
redshift data base provided by measurements of gamma-ray burst pair
annihilation lines \cite{liang86:a}. (Note the remarks 
in Sects.\ \ref{ssec:radii} and \ref{ssec:redshift} concerning the
interpretation of these data.)
From the redshift value of SNR N49 (if correct) we predict a neutron 
mass star of $1.1 \lsim M/\msun \lsim 1.6$, which is consistent with the 
observed mass range given in Sect.\ \ref{ssec:masses}.
The relativistic \eoss set a narrower mass limit for SNR N49 given by
$1.4 \lsim M/\msun \lsim 1.6$. 

Figure \ref{fig:ovrm1} displays the radius as a function of gravitational 
redshift. The solid dots refer to the maximum-mass star of  
each sequence. Of course, stars at their termination points
\begin{figure}[tb]
\begin{center}
\parbox[t]{6.5 cm}
{\leavevmode
\psfig{figure=rz.bb,width=6.0cm,height=7.0cm,angle=90}
{\caption[Radius versus gravitational redshift]{Radius as a function of
redshift for a sample of \eoss of Table \protect{\ref{tab:eos}}.}
\label{fig:ovrm1}}}
\ \hskip1.4cm   \
\parbox[t]{6.5cm}
{\leavevmode
\psfig{figure=im.bb,width=6.0cm,height=7.0cm,angle=90}
{\caption[Moment of inertia versus non-rotating mass]{Moment of inertia
as a function of mass for a sample of \eoss of
Table \protect{\ref{tab:eos}}.}\label{fig:ovimg1}}}
\end{center}
\end{figure}
possess the largest redshifts, and these become the smaller the lighter 
the stars. Under the assumption that the annihilation line interpretation 
is correct (Sect.\ \ref{ssec:redshift}), SNR N49 is predicted to 
have a radius in the range of 10 to 14 km.
The relativistic \eoss lead to a narrower radii range, i.e., 12.5 to 14 km.
In general, the non-relativistic \eoss lead to smaller radii for a given 
redshift. The reason for this lies in the relatively soft (stiff) behavior
at low (high) nuclear densities of the non-relativistic \eossp, which is less
pronounced for the relativistic Hartree and Hartree-Fock \eossp. 
(The softness/stiffness of the 
relativistic Brueckner-Hartree-Fock \eosp, however, is qualitatively 
similar to the one obtained for the non-relativistic models.) 
Small radius values of star models
are important in order to achieve rapid rotation. For that reason
star models constructed for the non-relativistic \eoss 
possess limiting rotational periods that are smaller than those obtained 
for the relativistic \eossp, as will be discussed in Sect.\ \ref{sec:rotst}). 
However, because of causality violation of the non-relativistic \eoss at
high nuclear densities (Sect.\ \ref{ssec:models}),
this trend may be an artifact. 

In Fig.\ \ref{fig:ovimg1} we show the moment
of inertia of neutron stars, given by \cite{glen92:crust}
\begin{eqnarray}
 I(\Omega) = 4\,\pi \,
\int\limits_0^{\pi/2} d\theta \int\limits_0^{R(\theta)} dr\,
e^{\lambda+\mu+\nu+\psi}\,
{{\epsilon + P(\epsilon)}\over{e^{2\nu-2\psi} - (\Omega-\omega)^2}} \,
{{\Omega-\omega}\over{\Omega}} \; ,
\label{eq:f231}
\end{eqnarray}
as a function of gravitational mass. In Sect.\ \ref{ssec:models} we have 
pointed out that, in general, the inclusion of baryon population in 
neutron star matter as well as the possible transition of confined 
hadronic matter to quark matter causes a softening of the \eosp,
which leads to somewhat smaller star masses and radii.
From the functional dependence of $I$ on radius and mass, which is 
of the form $I \propto R^2 \, M$, one expects a relative decrease of 
the \moi of star models constructed for such 
\eosp. Of course, the general relativistic expression for 
the \moip, given in Eq.\ (\ref{eq:f231}), is much more complicated. 
It accounts for the dragging effect of the local inertial 
frames (frequency dependence $\omega(r,\theta)$) and the curvature of 
space-time \cite{glen92:crust}. Nevertheless the qualitative 
dependence of $I$ on mass and radius as expressed in the classical
expression  remains valid \cite{arnett77:a}.
Estimates for the upper and lower bounds on the \moi of the Crab pulsar
derived from the pulsar's energy loss rate (labeled Rud72), and the lower 
bound on the moment of inertia derived from the luminosity of the Crab nebula
(labeled Crab) \cite{baym75:a,trimble70:a,borner73:a}
are shown in Fig.\ \ref{fig:ovimg1} for the purpose of comparison.
(The arrows refer only to the value of $I_{\rm Crab}$ and not to 
its mass, which is not known.)

\subsection{Rotating star models}\label{sec:rotst}

\subsubsection{Minimal rotational periods}\label{sssec:grav}

Figures \ref{fig:3} and \ref{fig:4} exhibit the limiting rotational 
periods of compact stars, which is set by the gravitational 
radiation reaction-driven instability 
\cite{lindblom86:a,lindblom92:a,weber90:e}. It originates
from counter-rotating surface vibrational modes, which at sufficiently
high rotational star frequencies are dragged forward.
In this case, gravitational radiation which inevitably accompanies
the aspherical transport of matter does not damp the modes, but
rather drives them \cite{chandrasekhar70:a,friedman83:a}.
Viscosity plays the important role of damping such gravitational-wave
radiation-reaction instabilities at a sufficiently reduced rotational
frequency such that the viscous damping rate and power in gravity
waves are comparable \cite{lindblom77:a}.
The instability modes are taken to have the dependence
${\rm exp}[i\omega_m(\Omega)t + i m\phi - t/\tau_m(\Omega)]$, where
$\omega_m$ is the frequency of the surface mode which depends on the
angular velocity $\Omega$ of the star, $\phi$ denotes the azimuthal angle,
and $\tau_m$ is the time scale for the mode which determines its growth
or damping.
The rotation frequency $\Omega$ at which it changes sign is the critical
frequency for the particular mode, $m$ (=2,3,4,...).  It is conveniently
expressed  as the frequency, denoted by $\Omega_m^\nu$, that solves
\cite{lindblom86:a} ($\nu$ refers to the viscosity dependence, see below)
\beqn
\Omega_m^\nu \, =\, {{\omega_m(0)}\over m } \, \left[
\tilde\alpha_m(\Omega_m^\nu) + \tilde\gamma_m(\Omega_m^\nu)\, \Bigl(
{ {\tau_{g,m}}\over{\tau_{\nu,m}} } \Bigr)^{1\over{2\,m + 1}}
\right]\; ,
\label{eq:grr1}
\eeqn
where
\beqn
\omega_m(0) \, \equiv \,
\sqrt{  { {2\,m\,(m-1)}\over{2\,m\,+\,1} } \, {M \over {R^3} }  }
\label{eq:omegm}
\eeqn
is the frequency of the vibrational mode in a non-rotating star.
The time scales for gravitational radiation-reaction
\cite{detweiler75:a}, $\tau_{g,m}$, and
for viscous damping time \cite{lamb81:a}, $\tau_{\nu,m}$, are
given by 
\beqn
\tau_{g,m} \, &=& \,
{2\over 3} \, { {(m-1)\, [(2m+1)!!]^2}\over{(m+1)\,(m+2)} } \,
\left( { {2m+1}\over{2 m (m-1)} } \right)^m \,
\left( { R \over M } \right)^{m+1} \, R \, ,
\label{eq:grr2} \\
\tau_{\nu,m} \, &=& \,
{ {R^2} \over {(2m+1)\, (m-1)} } ~ {1\over \nu} \; ,
\label{eq:grr3}
\eeqn
respectively.
The shear viscosity is denoted by $\nu$. It depends on the temperature,
$T$, of the star [$\nu(T) \propto T^{-2}$]. It is small in very hot
($T\approx 10^{10}$ K) and therefore young stars and larger in cold ones.
A characteristic feature of equations (\ref{eq:grr1})--(\ref{eq:grr3}) 
is that $\Omega^\nu_m$ merely depends on radius and mass
($R$ and $M$) of the spherical star model, which eases solving these
equations considerably. 

The functions $\tilde\alpha_m$ and $\tilde\gamma_m$
contain information about the pulsation of the rotating
star models and are difficult to determine
\cite{lindblom86:a,cutler87:a}. A reasonable first
step is to replace them by their corresponding Maclaurin
spheroid functions $\alpha_m$ and $\gamma_m$
\cite{lindblom86:a,cutler87:a}. We therefore take $\alpha_m(\Omega_m)$ and
$\gamma_m(\Omega_m)$ as calculated in Refs.\ \cite{ipser89:a,ipser90:a}
for the oscillations of rapidly rotating inhomogeneous Newtonian
stellar models (polytropic index $n$=1),
and Ref.\ \cite{lindblom86:a} for uniform-density Maclaurin spheroids
(i.e., $n$=0), respectively. Managan has shown that $\Omega^\nu_m$ depends
much more strongly on the \eos and the mass of the neutron star
model [through $\omega_m(0)$ and $\tau_{g,m}$, see Eqs.\ (\ref{eq:omegm}] and
(\ref{eq:grr2})) than on the polytropic index assumed in calculating
$\alpha_m$ \cite{managan86:a}.

Figure \ref{fig:3} shows the critical rotational periods, at which
emission of gravity waves sets in in hot ($T = 10^{10}$ K) 
pulsars, newly born in supernova explosions.
Figure \ref{fig:4} is the analog of Fig.\ \ref{fig:3}, but for
old and therefore cold compact stars of temperature $T=10^6$ K, like
neutron stars in binary systems that are being spun up (and thereby reheated)
by mass accretion from a companion.
One sees that the limiting rotational 
periods $P^T$ ($\equiv 2\pi/\Omega^\nu_m$)
are the smaller the more massive (and thus the smaller the radius) 
the star model (cf.\ Fig.\ \ref{fig:ovrm1}). 
A comparison between Figs.\ \ref{fig:3} and 
\ref{fig:4} shows that the instability periods are shifted
toward smaller values the colder the star, due to the larger viscosity
in such objects. 
Consequently, the instability modes of compact stars in binary systems are 
excited at smaller rotational periods than is the case for hot and newly
born pulsars in supernovae.\footnote{Sawyer \protect{\cite{sawyer89:b}}
has found that the bulk viscosity of neutron star matter goes as 
the sixth power of the temperature, as compared with a $T^{-2}$ dependence 
for the shear viscosity which is treated here. This means that at 
$T \gsim 10^9$ K the bulk viscosity would dominate over the shear viscosity
and thus damp the gravitational-wave instability.
In this case the instability periods would be shifted toward
values that are relatively close to the Kepler period $\pkgr$
\protect{\cite{ipser90:a,ipser91:a}}, which will be discussed below.
The latter sets an absolute 
limit on stable rotation because of mass shedding. 
Bounds on $\pkgr$ are given in Table \protect{\ref{tab:lu}}.} 
The dependence of $P^T$ on the \eos is shown too in these figures. One 
\begin{figure}[tb]
\begin{center}
\parbox[t]{6.5 cm}
{\leavevmode
\psfig{figure=grr10.bb,width=6.0cm,height=7.0cm,angle=90}
{\caption[Limiting rotational periods versus mass for hot, newly born
stars]{Gravitational radiation-reaction instability period $P^T$ versus 
mass for newly born stars of temperature $T = 10^{10}$ K 
\protect{\cite{weber91:c}}.}\label{fig:3}}}
\ \hskip1.4cm   \
\parbox[t]{6.5cm}
{\leavevmode
\psfig{figure=grr06.bb,width=6.0cm,height=7.0cm,angle=90}
{\caption[Limiting rotational periods versus mass for cold and therefore old
stars]{Gravitational radiation-reaction instability period $P^T$ versus 
mass for old stars of temperature $T = 10^6$ K \protect{\cite{weber91:c}}.}
\label{fig:4}}}
\end{center}
\end{figure}
sees that the lower limits on $P^T$ are set by the non-relativistic \eos 
labeled 16 due to the small radii values obtained for the star models 
constructed from it. The relativistic models for the \eos
generally lead to larger rotational periods due to the somewhat larger
radii of the associated star models.

The rectangles in Figs.\ \ref{fig:3} and \ref{fig:4} denoted
``observed'' cover both the range of observed neutron star masses,
$1.1\lsim M/\msun \lsim 1.8$ as well as observed pulsar periods, i.e.,
$P\geq 1.6$ ms.  One sees that even the most rapidly rotating pulsars
so far observed have rotational periods larger that those at which
gravity-wave emission sets in and thus can be understood as rotating
neutron or hybrid stars.\footnote{Depending on their composition,
  compact star models are denoted as neutron, hybrid, or strange
  stars.  Neutron stars consist of protons, neutrons and more massive
  baryons in $\beta$ equilibrium with leptons; hybrid stars are
  compact stars which, in addition to baryons, also contain quarks in
  their dense cores; hypothetical strange stars consist of (3-flavor)
  strange quark matter which, by hypothesis, may form the absolute
  ground-state of strongly interacting matter (see Sect.\ 
  \ref{sec:strange} for more details).} The observation of pulsars
possessing masses in the observed range but rotational periods that
are smaller than say $\sim 1$ ms (depending on temperature and thus on
the pulsar's history) would be in clear contradiction to our \eossp.
Consequently the possible future observation of such pulsars cannot be
reconciled -- in the framework of our collection of models for
superdense neutron star matter -- with the interpretation of such
objects as rapidly rotating neutron or hybrid stars.  This conclusion
is strengthened by the construction of neutron star models that are
rotating at their Kepler periods, $\pkgr$, at which mass shedding at
the star's equator sets in. Therefore this period sets an absolute
limit on rapid rotation, which cannot be overcome by any rapidly
rotating star.  On the basis of neutron star models constructed from
the selection of \eoss studied here, the smallest Kepler periods are
found to be in the range of $0.7\lsim\pkgr\lsim 1$~ms, depending on
the softness of the \eosp\footnote{An investigation of this period for
  neutron stars and other compact objects that is performed without
  taking recourse to any particular models of dense matter (but
  derives the limit only on the general principles that: Einstein's
  equations describe stellar structure, matter is microscopically
  stable, and causality is not violated) has only recently been
  performed by Glendenning \cite{glen92:limit}.  He establishes a
  lower bound for the minimum Kepler period for a $M=1.442\, \msun$
  neutron star of $\pkgr= 0.33$ ms.  Of course the \eos that nature
  has chosen need not be the one that allows stars to rotate most
  rapidly.} (cf.\ Table \ref{tab:lu}). $\pkgr$ is given by
\cite{friedman86:a,weber91:d,glen93:drag} \beqn \pkgr \equiv \frac{2\,
  \pi}{\okgr} \, , ~{\rm with} ~~ \okgr = \omega
+\frac{\omega^\prime}{2\psi^\prime} + e^{\nu -\psi} \sqrt{
  \frac{\nu^\prime}{\psi^\prime} + \Bigl(\frac{\omega^\prime}{2
    \psi^\prime}e^{\psi-\nu}\Bigr)^2} \; .
  \label{eq:okgr}
  \eeqn
Note that $\pkgr$ can only be obtained by means of solving
Eq.\ (\ref{eq:okgr}) self-consistently in combination with Einstein's equation,
\beqn 
 {\cal R}^{\kappa\lambda} \; - \; {1\over 2} \; g^{\kappa\lambda} \; {\cal R}
 \; = \; 8\, \pi \; {\cal T}^{\kappa\lambda}(\epsilon,P(\epsilon)) \;\; ,
 \label{eq:einstein}
 \eeqn and the equation of energy-momentum conservation, ${\cal
   T}{^{\kappa\lambda}}_{;\lambda}=0$, which renders the problem
 extremely complicated.  The quantities ${\cal R}^{\kappa\lambda}$,
 $g^{\kappa\lambda}$, and ${\cal R}$ denote the Ricci tensor, metric
 tensor, and Ricci scalar (scalar curvature), respectively.  The
 dependence of the energy-momentum tensor ${\cal T}^{\kappa\lambda}$
 on pressure and energy density, $P$ and $\epsilon$ respectively, is
 indicated in Eq.\ (\ref{eq:einstein}).  The quantities $\omega$,
 $\nu$, and $\psi$ in Eq.\ (\ref{eq:okgr}) denote the frame dragging
 frequency of local inertial frames, and time- and space-like metric
 functions, respectively.

\subsubsection{Bounds on properties of rapidly rotating pulsars}
\label{sssec:bounds}

We restrict ourselves to the properties of 
a rapidly rotating pulsar model having a mass of $M\approx 1.45\,\msun$, as 
supported by the evolutionary history of supermassive stars \cite{shapiro83:a}.
The bounds on its properties, whose knowledge is of great importance for 
the interpretation of fast pulsar, are summarized in Table \ref{tab:lu}.
The listed properties are:
period at which the gravitational radiation-reaction instability sets in,
$P^T$ (in ms) with star temperature listed in parentheses;
Kepler period, $\pkgr$ (in ms);
central energy density, $\epsc$ 
(in units of the density of normal nuclear matter);
\moip, $I$ (in $\gcmsq$);
redshifts of photons emitted at the star's equator in backward ($\zbw$)
and forward ($\zfw$) direction, defined by
\begin{eqnarray}
  z_{\rm B\,/\,F} (\Omega) =  e^{-\nu(\Omega)} \;
       \left( 1 \pm \omega(\Omega)\, e^{\psi(\Omega)-\nu(\Omega)}
     \right)^{-1} \; 
     \left( {{1\pm V(\Omega)}\over{1\mp V(\Omega)}} \right)^{1/2}
     \; - \; 1 \; ,
\label{eq:f273}
\end{eqnarray}
and the redshift of photons emitted from the star's pole,
\begin{eqnarray}
    z_{\rm p} (\Omega) \quad = \quad e^{-\,\nu(\Omega)} \quad - \quad 1\; .
\label{eq:f274p}
\end{eqnarray} 
($V$ denotes the star's velocity at the equator \cite{friedman86:a,weber91:d}.
All other quantities are as in Sect.\ \ref{sssec:grav}.)
According to Table \ref{tab:lu}, newly born 
pulsars observed in supernova explosions can only rotate stably
at periods $\gsim 1$ ms.
Half-millisecond periods, for example, are completely excluded
for pulsars made of baryon matter. Therefore, the possible
future discovery of a single sub-millisecond pulsar, rotating with a 
period of say $\sim 0.5$ ms,
would give a strong hint that such an object is a rotating strange star,
not a neutron star, and that 3-flavor strange quark matter
may be the true ground-state of the strong interaction, as pointed 
out by Glendenning \cite{glen91:a}.
Old pulsar of $T=10^6$ K (and mass $M\approx 1.45\, \msun$) cannot be
spun up to stable rotational periods smaller than $\approx 0.8$ ms.
\begin{table}[tb]
\caption[Theoretically determined lower and upper bounds on the properties
of a rotating neutron star of $M\approx 1.45 \,\msun$]{Lower and upper
bounds on the properties of pulsars with $M\approx 1.45 \,\msun$, 
calculated for the broad collection of \eoss of Table \ref{tab:eos} 
\protect{\cite{weber92:a}}.
\label{tab:lu}}
\begin{center}
\begin{tabular}{ccccccccc} \hline
    &$P^T\,(10^6\;{\rm K})$   &$P^T\,(10^{10}\;{\rm K})$    &$\pkgr$
&$\epsc/\epsilon_0$   &log $I$     &$\zbw$   &$\zfw$   &$\zpole$  \\
\hline \hline \\
upper bound  &1.1    &1.5    &1    &5    &45.19    &1.05   &$-0.18$  &0.45  \\
lower bound  &0.8    &1.1    &0.7  &2    &44.95    &0.59   &$-0.21$  &0.23  \\
\hline
\end{tabular}
\end{center}
\end{table}
Again, the two fastest yet observed pulsars, rotating at 1.6 ms,
are compatible with the periods in Table \ref{tab:lu}, provided their masses
are larger than 1 $\msun$ \cite{weber91:c}. For the purpose of comparison
the Kepler period, below which mass shedding at the star's equator sets in,
is listed too. It may play a role in cold (hot) pulsar whose 
rotation is stabilized by its large shear (bulk) viscosity value (see footnote
4 on Sawyers calculation of the viscosity in dense nuclear matter).

\section{Strange quark matter stars}\label{sec:strange}

\subsection{The strange matter hypothesis}\label{ssec:hyp}

The hypothesis that strange quark matter may be the absolute ground state
of the strong interaction (not $^{56}{\rm Fe}$)
has been raised by Bodmer \cite{bodmer71:a}, Witten \cite{witten84:a},
and Terazawa \cite{terazawa89:a}.
If the hypothesis is true, then a separate class of compact stars could
exist, which are called strange stars. They form a distinct and disconnected
branch of compact stars and are not part of the continuum of equilibrium
configurations that include white dwarfs and neutron stars. In principle
both strange and neutron stars could exist. However if strange stars exist,
the galaxy is likely to be contaminated by strange quark nuggets which
would convert all neutron stars to strange
stars \cite{glen91:a,madsen91:a,caldwell91:a}.
This in turn means that the objects known to astronomers as pulsars
are probably rotating strange matter stars, not neutron matter stars
as is usually assumed.
Unfortunately the bulk properties of models of neutron and strange stars
of masses that are typical for neutron stars,
$1.1 \lsim M/\msun \lsim 1.8$, are relatively similar
and therefore do not allow the distinction between the two possible pictures.
The situation changes however as regards the possibility of 
 fast rotation of strange stars. This has its origin
in the completely different mass-radius relations
of neutron and strange stars (see Fig.\ \ref{fig:radss}) \cite{glen89:j}.
As a consequence of this the entire family of strange stars can rotate
rapidly -- i.e., considerably below one millisecond --, 
not just those near the limit of gravitational
collapse to a black hole as is the case for neutron stars. 
Furthermore, as I shall discuss below, the cooling history of 
neutron stars and strange stars may be quite different.  

\subsection{Hadronic crust on strange stars and pulsar glitches}
\label{ssec:crust}

At the present time there appears to be only one crucial astrophysical 
test of the strange-quark-matter hypothesis, and that is whether 
strange quark stars can give rise to the observed phenomena of pulsar
glitches. In the crust quake model
an oblate solid nuclear crust in its present shape slowly comes out of
equilibrium with the forces acting on it as the rotational period changes,
and fractures when the built up stress exceeds the sheer strength of the 
crust material. The period and rate of change of period slowly heal to 
the trend preceding the glitch as the coupling between crust and core
re-establish their co-rotation. The existence of glitches may have a decisive
impact on the question of whether strange matter is the ground state
of the strong interaction.  

The only existing investigation which deals with the calculation of the 
thickness, mass and moment of inertia of the nuclear solid crust that can 
exist on the surface of a rotating, general relativistic strange quark 
star has only recently been performed by Glendenning and Weber
\cite{glen92:crust}. Their calculated mass-radius relationship 
for strange stars with a nuclear crust, whose maximum density is the
neutron drip density, is shown in Fig.\ \ref{fig:radss}.
(Free neutrons in the star cannot exist. These would be dissolved into 
quark matter as they gravitate into the strange core. 
Therefore the maximum density 
\begin{figure}[tb]
\begin{center}
\parbox[t]{6.5cm}
{\leavevmode
\psfig{figure=rm.bb,width=6.0cm,height=7.0cm,angle=90}
{\caption[Radius as a function of mass of a strange star 
with crust, and radius of the strange star core for inner crust density
equal to neutron drip, for non-rotating stars. The bag constant is
$\bag=160$ MeV. The solid dots refer to the maximum-mass model of the 
sequence]{Radius as a function of mass of a non-rotating strange star 
with crust \protect{\cite{glen92:crust}}.}\label{fig:radss}}}
\ \hskip1.4cm \
\parbox[t]{6.5cm}
{\leavevmode
\psfig{figure=icit.bb,width=6.0cm,height=7.0cm,angle=90}
{\caption[The ratio $\icrust/\itotal$ as a function of star mass. Rotational 
frequencies are shown as a fraction of the Kepler frequency. The solid 
dots refer to the maximum-mass models. The bag constant is 
$\bag=160$ MeV]{The ratio $\icrust/\itotal$ as a function of star mass. 
Rotational frequencies are shown as a fraction of the Kepler frequency,
$\okgr$ \protect{\cite{glen91:c}}.}
\label{fig:cm160}}}
\end{center}
\end{figure}
of the crust is strictly limited by neutron drip. This density is
about $4.3\times 10^{11}~\gcmt$.)
Since the crust is bound by the gravitational interaction (and not
by confinement, which is the case for the strange matter core), the 
relationship is qualitatively similar to the one for neutron and hybrid
stars, as can be seen from Fig.\ \ref{fig:ovrm1}. The radius being largest
for the lightest and smallest for the heaviest stars 
(indicated by the soid dot in Fig.\ \ref{fig:radss}) in the sequence.
Just as for neutron stars the relationship is not necessarily 
monotonic at intermediate masses. The radius of the strange quark core,
denoted $\rdrip$, is shown by the dashed line. (A value for the bag constant 
of $\bag=160$ MeV for which 
3-flavor strange matter is absolutely 
stable has been chosen. This choice represents
weakly bound strange matter with an energy per baryon $\sim 920$ MeV,
and thus corresponds to strange quark matter being
absolutely bound with respect to $^{56}{\rm Fe}$).
The radius of the strange quark core is proportional to $M^{1/3}$ which is 
typical for self-bound objects. This proportionality is only modified near 
that stellar mass where gravity terminates the stable sequence.
The sequence of stars has a minimum mass of $\sim 0.015\, \msun$ (radius of
$\sim 400$ km) or about 15 Jupiter masses, which is considerably 
smaller than that of 
neutron star sequences, about $0.1~\msun$ \cite{baym71:a}.
The low-mass strange stars may be of considerable importance since they
may be difficult to detect and therefore may effectively hide 
baryonic matter.
Furthermore, of interest to the subject of cooling of strange stars is the 
crust thickness of strange stars \cite{pizzochero91:a}. 
It ranges from $\sim~ 400$ km 
for stars at the lower mass limit to $\sim 12$ km for stars of mass
$\sim 0.02\,\msun$, and is a fraction of a kilometer for the star at
the maximum mass \cite{glen92:crust}. 

The moment of inertia of the hadronic crust, $\icrust$, that can be carried
by a strange star as a function of star mass for a sample of rotational 
frequencies of $\Omega=\okgr,\okgr/2$ and 0 is shown in Fig.\
\ref{fig:cm160}. Because of the relatively small crust mass of the
maximum-mass models of each sequence, the ratio
$\icrust/\itotal$ is smallest for them (solid dots in Fig.\
\ref{fig:cm160}). The less massive the strange star the larger its radius
(Fig.\ \ref{fig:radss}) and therefore the larger both $\icrust$ as well as 
$\itotal$. The dependence of $\icrust$ and $\itotal$ on $M$ is such that 
their ratio $\icrust/\itotal$ is a monotonically decreasing function of $M$. 
One sees that there is only a slight difference between $\icrust$ for 
$\Omega=0$ and $\Omega=\okgr/2$.

Of considerable relevance for the question of whether strange stars can 
exhibit glitches in rotation frequency, one sees that $\icrust/\itotal$ 
varies between $10^{-3}$ and $\sim 10^{-5}$ at the maximum mass.
If the angular momentum of the pulsar is conserved in the quake
then the relative frequency change and moment of inertia change are equal,
and one arrives at \cite{glen92:crust}
\beqn
       {{\Delta \Omega}\over{\Omega}} \; = \; 
       {{|\Delta I|}\over {I_0}} \; > \;
       {{|\Delta I|}\over {I}} \; \equiv \; f \;
       {\icrust\over I}\; \sim \; (10^{-5} - 10^{-3})\, f \; , 
       ~{\rm with}  \quad 0 < f < 1\; .
\label{eq:delomeg}
\eeqn
Here $I_0$ denotes the moment of inertia of that part of the star whose
frequency is changed in the quake. It might be that of the crust only, or some
fraction, or all of the star. The factor $f$ in Eq.\ (\ref{eq:delomeg})
represents the fraction of the crustal moment of inertia that is altered in 
the quake, i.e., $f \equiv |\Delta I|/ \icrust$. 
Since the observed glitches have
relative frequency changes $\Delta \Omega/\Omega = (10^{-9} - 10^{-6})$,
a change in the crustal moment of inertia of $f\lsim 0.1$ would cause
a giant glitch even in the least favorable case (for more details, see
\cite{glen92:crust}). Moreover,   
we find that the observed range of the fractional change in
the spin-down rate,
$\dot \Omega$, is consistent with the crust having the small moment of inertia
calculated and the quake involving only a small fraction $f$ of that,
just as in Eq.\ (\ref{eq:delomeg}).
For this purpose we write \cite{glen92:crust}
\beqn
        { {\Delta \dot\Omega}\over{\dot\Omega } } \; = \;
        { {\Delta \dot\Omega /  \dot\Omega} \over  
          {\Delta    \Omega  /     \Omega }  } \,
        { {|\Delta I |}\over{I_0} } \; = \; 
        { {\Delta \dot\Omega /  \dot\Omega} \over  
          {\Delta    \Omega  /     \Omega }  } \; f \;
         {\icrust\over {I_0} } \; > \; 
       (10^{-1}\; {\rm to} \; 10) \; f \; ,
\label{eq:omdot}
\eeqn
where use of Eq.\ (\ref{eq:delomeg}) has been made. Equation (\ref{eq:omdot})
 yields a small $f$ value, i.e.,  
$f < (10^{-4} \; {\rm to} \; 10^{-1})$, in agreement with $f\lsim 10^{-1}$
established just above. Here measured values of the ratio 
$(\Delta \Omega/\Omega)/(\Delta\dot\Omega/\dot\Omega) \sim 10^{-6}$ to 
$10^{-4}$ for the Crab and Vela pulsars, respectively, have been used.

\subsection{Thermal evolution of neutron and strange stars}
\label{ssec:thermal}

\begin{figure}[tb]
\begin{center} 
\leavevmode
\psfig{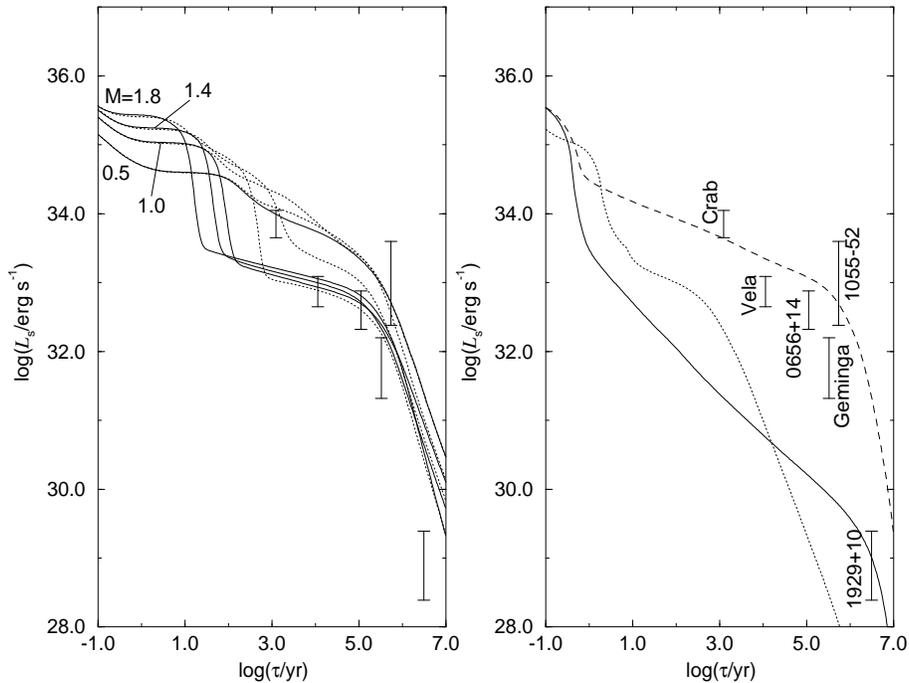} 
\caption{Left panel: Cooling of neutron stars with pion (solid
curves) or kaon condensates (dotted curve). Right panel: Cooling of
$M=1.8\,M_\odot$ strange stars with crust. The cooling curves of
  lighter strange stars, e.g. $M\gsim 1\,M_\odot$, differ only
  insignificantly from those shown here. Three different assumptions
  about a possible superfluid behavior of strange quark matter are
  made: no superfluidity (solid), superfluidity of all three flavors
  (dotted), and superfluidity of up and down flavors only
  (dashed). The vertical bars denote luminosities of observed pulsars.
  \label{fig:cool}}
\end{center}
\end{figure}

The left panel of Fig. \ref{fig:cool} shows a numerical simulation of
the thermal evolution of neutron stars.  The neutrino emission rates
are determined by the modified and direct Urca processes, and the
presence of a pion or kaon condensate.  The baryons are treated as
superfluid particles. Hence the neutrino emissivities are suppressed
by an exponential factor of $\exp(-\Delta/kT)$, where $\Delta$ is the
width of the superfluid gap (see Ref.\, \cite{schaab95:a} for
details).  Due to the dependence of the direct Urca process and the
onset of meson condensation on star mass, stars that are too light for
these processes to occur (i.e., $M<1\,\msun$) are restricted to
standard cooling via modified Urca.  Enhanced cooling via the other
three processes results
in a sudden drop of the star's surface temperature after about 10 to
$10^3$ years after birth, depending on the thickness of the ionic
crust.  As one sees, agreement with the observed data is achieved
only if different masses for the underlying pulsars are assumed.
The right panel of Fig. \ref{fig:cool} shows cooling simulations of
strange quark stars. The curves differ with respect to assumptions
made about a possible superfluid behavior of the quarks. Because of
the higher neutrino emission rate in non-superfluid quark matter, such
quark stars cool most rapidly (as long as cooling is core
dominated). In this case one does not get agreement with most of the
observed pulsar data. The only exception is pulsar PSR
1929+10. Superfluidity among the quarks reduces the neutrino emission
rate, which delays cooling\,\cite{schaab95:a}. This moves the cooling
curves into the region where most of the observed data lie.

Subject to the inherent uncertainties in the behavior of strange quark
matter as well as superdense nuclear matter, at present it appears
much too premature to draw any definitive conclusions about the true
nature of observed pulsars. Nevertheless, should a continued future
analysis in fact confirm a considerably faster cooling of strange
stars relative to neutron stars, this would provide a definitive
signature (together with rapid rotation) for the identification of a
strange star. Specifically, the prompt drop in temperature at the very
early stages of a pulsar, say within the first 10 to 50 years after
its formation, could offer a good signature of strange
stars\,\cite{pizzochero91:a}.  This feature, provided it withstands a
more rigorous analysis of the microscopic properties of quark matter,
could become particularly interesting if continued observation of SN
1987A would reveal the temperature of the possibly existing pulsar at
its center.

\section{Summary}

This work begins with an investigation of the properties of superdense
neutron star matter. Various models for the associated \eos of such
matter are introduced and their characteristic features are discussed
in great detail.  These \eoss are then applied to the construction of
models of non-rotating as well as rotating compact stars, and an
investigation of their cooling behavior.  The theoretically computed
neutron star properties are compared with the body of observed data.

The indication of this work is that the gravitational radiation-reaction
driven instability in neutron stars 
sets a lower limit on stable rotation for massive 
 neutron stars of $P \approx$ 0.8 ms. 
Lighter ones having typical pulsar masses of 
$1.45\, \msun$ are predicted to have stable rotational periods $P\gsim 1$ ms. 
This finding may have very important implications for the nature of any
pulsar that is found to have a shorter period, say around $P\approx 0.5$ ms.
Since our representative collection of nuclear \eoss does not allow for
rotation at such small periods, the interpretation of such objects as
rapidly rotating neutron stars would fail. 
Such objects, however, can be understood as rapidly rotating 
 strange stars. The plausible
ground-state state in that event is the deconfined phase of (3-flavor)
strange quark matter. From the QCD energy scale this is as likely a
ground-state as the confined phase of atomic matter.
At the present time there appears to be only one 
crucial astrophysical test of the strange quark matter hypothesis, and that 
is whether strange quark stars can give rise to the observed phenomena of 
pulsar glitches. We demonstrate that the nuclear solid crust that can
exist on the surface of a strange star can have a moment of inertia
sufficiently large that a fractional change can account for the 
magnitude of pulsar glitches. Furthermore low-mass strange 
stars can have enormously large nuclear crusts (up to several thousand
kilometers \cite{weber93:b,glen94:a}) 
which enables such objects to be possible hiding places of baryonic matter. 
Finally, due to the uncertainties in the behavior of superdense nuclear
as well as strange matter, we conclude that no definitive conclusions
about the true nature (strange or conventional) of observed pulsars can
be drawn from cooling simulations yet. As of yet, they could be made of
strange quark matter as well as of conventional nuclear matter.

\bigskip
{\bf Acknowledgement:}
\doe
\vskip 2.truecm


\clearpage


\baselineskip=10.pt

\begin{thebibliography}{100}

\bibitem{manchester77:a}
R. N. Manchester and J. H. Taylor, {\it Pulsars}, W. H. Freeman and Co., San
  Francisco, 1977.

\bibitem{backer82:a}
D. C. Backer, S. R. Kulkarni, C. Heiles, M. M. Davis, and W. M. Goss, Nature
  {\bf 300} (1982) 615.

\bibitem{manchester91:a}
R. N. Manchester, A. G. Lyne, C. Robinson, N. D'Amico, M. Bailes, and J. Lim,
  Nature {\bf 352} (1991) 219.

\bibitem{friedman86:a}
J. L. Friedman, J. R. Ipser, and L. Parker, Astrophys.\ J.\ {\bf 304} (1986)
  115.

\bibitem{friedman89:a}
J. L. Friedman, J. R. Ipser, and L. Parker, Phys.\ Rev.\ Lett.\ {\bf 62} (1989)
  3015.

\bibitem{lattimer90:a}
J. M. Lattimer, M. Prakash, D. Masak, and A. Yahil, Astrophys.\ J.\ {\bf 355}
  (1990) 241.

\bibitem{weber90:a}
F. Weber, N. K. Glendenning, and M. K. Weigel, Astrophys.\ J.\ {\bf 373} (1991)
  579.

\bibitem{weber91:b}
F. Weber and N. K. Glendenning, Astrophys.\ J.\ {\bf 390} (1992) 541.

\bibitem{weber91:d}
F. Weber and N. K. Glendenning, {\it Hadronic Matter and Rotating Relativistic
  Neutron Stars}, Proceedings of the Nankai Summer School, ``Astrophysics and
  Neutrino Physics'', p. 64--183, Tianjin, China, 17-27 June 1991, ed. by D. H.
  Feng, G. Z. He, and X. Q. Li, World Scientific, Singapore, 1993.

\bibitem{witten84:a}
E. Witten, Phys.\ Rev.\ D {\bf 30} (1984) 272.

\bibitem{alcock86:a}
C. Alcock, E. Farhi, and A. V. Olinto, Astrophys.\ J.\ {\bf 310} (1986) 261.

\bibitem{haensel86:a}
P. Haensel, J. L. Zdunik, and R. Schaeffer, Astron.\ Astrophys.\ {\bf 160}
  (1986) 121.

\bibitem{sprung72:a}
D. W. L. Sprung, Adv.\ Nucl.\ Phys.\ {\bf 5} (1972) 225.

\bibitem{day78:a}
B. D. Day, Rev.\ Mod.\ Phys.\ {\bf 51} (1979) 821.

\bibitem{pandharipande79:a}
V. R. Pandharipande and R. B. Wiringa, Rev.\ Mod.\ Phys.\ {\bf 51} (1979) 821.

\bibitem{wiringa88:a}
R. B. Wiringa, V. Fiks, and A. Fabrocini, Phys.\ Rev.\ C {\bf 38} (1988) 1010.

\bibitem{glen85:b}
N. K. Glendenning, Astrophys.\ J.\ {\bf 293} (1985) 470.

\bibitem{holinde72:a}
K. Holinde, K. Erkelenz, and R. Alzetta, Nucl.\ Phys.\ {\bf A194} (1972) 161;
  {\bf A198} (1972) 598.

\bibitem{machleidt87:a}
R. Machleidt, K. Holinde, and Ch. Elster, Phys.\ Rep.\ {\bf 149} (1987) 1.

\bibitem{glen85:a}
N. K. Glendenning, Phys.\ Lett.\ {\bf 114B} (1982) 392; \\ N. K. Glendenning,
  Astrophys.\ J.\ {\bf 293} (1985) 470; \\ N. K. Glendenning, Z.\ Phys.\ A {\bf
  326} (1987) 57; \\ N. K. Glendenning, Z.\ Phys.\ A {\bf 327} (1987) 295.

\bibitem{weber89:e}
F. Weber and M. K. Weigel, Nucl.\ Phys.\ {\bf A505} (1989) 779.

\bibitem{martin59:a}
P. C. Martin and J. Schwinger, Phys.\ Rev.\ {\bf 115} (1959) 1342.

\bibitem{wilets79:a}
L. Wilets, Green's functions method for the relativistic field theory many-body
  problem, in {\it Mesons in nuclei}, Vol. III, ed. by M. Rho, D. Wilkinson,
  North-Holland, Amsterdam, 1979.

\bibitem{huber93:a}
H. Huber, F. Weber, and M. K. Weigel, Phys.\ Lett.\ {\bf 317B} (1993) 485.

\bibitem{huber94:a}
H. Huber, F. Weber, and M. K. Weigel, Phys.\ Rev.\ C {\bf 50} (1994) R1287.

\bibitem{huber95:a}
H. Huber, F. Weber, and M. K. Weigel, Phys.\ Rev.\ C {\bf 51} (1995) 1790.

\bibitem{poschenrieder88:a}
P. Poschenrieder and M. K. Weigel, Phys.\ Lett.\ {\bf 200B} (1988) 231; \\ P.
  Poschenrieder and M. K. Weigel, Phys.\ Rev.\ C {\bf 38} (1988) 471.

\bibitem{weber88}
F. Weber and M. K. Weigel, Z.\ Phys.\ A{\bf 330} (1988) 249.

\bibitem{weber89:a}
F. Weber and M. K. Weigel, J.\ Phys.\ G {\bf 15} (1989) 765.

\bibitem{serot86:a}
B. D. Serot and J. D. Walecka, Adv.\ Nucl.\ Phys.\ {\bf 16} (1986) 1.

\bibitem{horowitz87:a}
C. J. Horowitz and B. D. Serot, Nucl.\ Phys.\ {\bf A464} (1987) 613.

\bibitem{glen89:a493}
N. K. Glendenning, Nucl.\ Phys.\ {\bf A493} (1989) 521.

\bibitem{glen91:c}
N. K. Glendenning, Nucl.\ Phys.\ B (Proc. Suppl.) {\bf 24B} (1991) 110.

\bibitem{glen91:pt}
N. K. Glendenning, Phys.\ Rev.\ D {\bf 46} (1992) 1274.

\bibitem{glen91:b}
N. K. Glendenning, F. Weber, and S. A. Moszkowski, Phys.\ Rev.\ C {\bf 45}
  (1992) 844.

\bibitem{glen86:b}
N. K. Glendenning, Phys.\ Rev.\ Lett.\ {\bf 57} (1986) 1120.

\bibitem{bethe74:a}
H. A. Bethe and M. Johnson, Nucl.\ Phys.\ {\bf A230} (1974) 1.

\bibitem{friedman81:a}
B. Friedman and V. R. Pandharipande, Nucl.\ Phys.\ {\bf A361} (1981) 502.

\bibitem{myers90:a}
W. D. Myers and W. J. Swiatecki, Ann.\ Phys.\ (N.\ Y.) {\bf 204} (1990) 401.

\bibitem{myers91:a}
W. D. Myers and W. J. Swiatecki, Ann.\ Phys.\ (N.\ Y.) {\bf 211} (1991) 292.

\bibitem{myers94:a}
W. D. Myers and W. J. Swiatecki, ``The Nuclear Thomas-Fermi Model'', presented
  by W. J. Swiatecki at the XXIX Zakopane School of Physics, September 5--14,
  1994, Zakopane, Poland, (LBL-36004).

\bibitem{myers94:c}
W. D. Myers and W. J. Swiatecki, ``Thomas Fermi Treatment of Nuclear Masses,
  Deformations and Density Distributions'', presented by W. J. Swiatecki at the
  Fourth KINR International School on Nuclear Physics, Kiev, Russia, August
  29--September 7, 1994, Zakopane, Poland, (LBL-36005).

\bibitem{pandharipande71:a}
V. R. Pandharipande, Nucl.\ Phys.\ {\bf A178} (1971) 123.

\bibitem{schiavilla86:a}
V. R. Pandharipande and R. B. Wiringa, Nucl.\ Phys.\ {\bf A449} (1986) 219.

\bibitem{wiringa84:a}
R. B. Wiringa, R.A. Smith, and T. L. Ainsworth, Phys.\ Rev.\ C {\bf 29} (1984)
  1207.

\bibitem{lagaris81:a}
I. E. Lagaris and V. R. Pandharipande, Nucl.\ Phys.\ {\bf A359} (1981) 349.

\bibitem{myers69:a}
W. D. Myers, {\it Droplet Model of Atomic Nuclei} (New York: McGraw Hill,
  1977);\ \ W. D. Myers and W. Swiatecki, Ann.\ of Phys.\ {\bf 55} (1969)
  395;\\ W. D. Myers and K.-H. Schmidt, Nucl.\ Phys.\ {\bf A410} (1983) 61.

\bibitem{banj71:a}
P. K. Banerjee and D. W. L. Sprung, Can.\ J.\ Phys.\ {\bf 49} (1971) 1871.

\bibitem{mahaux81:a}
C. Mahaux, {\it Brueckner Theory of Infinite Fermi Systems}, Lecture Notes in
  Physics, Vol. 138, (Springer Verlag, Berlin, 1981).

\bibitem{fiset72:a}
E. O. Fiset and T. C. Foster, Nucl.\ Phys.\ {\bf A184} (1972) 588.

\bibitem{kim74:a}
Q. Ho-Kim and F. C. Khanna, Ann.\ Phys.\ {\bf 86} (1974) 233.

\bibitem{weber85}
F. Weber and M. K. Weigel, Phys.\ Rev.\ C {\bf 32} (1985) 2141.

\bibitem{coester70:a}
F. Coester, S. Cohen, B. Day, and C. M. Vincent, Phys.\ Rev.\ C {\bf 1} (1970)
  769.

\bibitem{wong72:a}
C. W. Wong, Ann.\ Phys.\ (N. Y.)\ {\bf 72} (1972) 107.

\bibitem{celenza81:a}
L. S. Celenza and C. M. Shakin, Phys.\ Rev.\ C {\bf 24} (1981) 2704.

\bibitem{celenza86:a}
L. S. Celenza and C. M. Shakin, {\it Relativistic Nuclear Structure Physics},
  World Scientific Lecture Notes in Physics, Vol. 2, World Scientific,
  Singapore, 1986.

\bibitem{malfliet87:a}
B. ter Haar and R. Malfliet, Phys.\ Rev.\ Lett.\ {\bf 59} (1987) 1652.

\bibitem{weber91:a}
F. Weber and N. K. Glendenning, Phys.\ Lett.\ {\bf 265B} (1991) 1.

\bibitem{glen89:j}
N. K. Glendenning, {\it Supernovae, Compact Stars and Nuclear Physics}, invited
  paper in Proc.\ of 1989 Int. Nucl. Phys. Conf., Sao Paulo, Brasil, Vol. 2,
  ed. by M. S. Hussein et al., World Scientific, Singapore, 1990.

\bibitem{glen89:k}
N. K. Glendenning, {\it Strange-Quark-Matter Stars}, Proc. Int. Workshop on
  Relativ. Aspects of Nucl. Phys., Rio de Janeiro, Brasil, 1989, ed. by T.
  Kodama, K. C. Chung, S. J. B. Duarte, and M. C. Nemes, World Scientific,
  Singapore, 1990.

\bibitem{ellis91:a}
J. Ellis, J. I. Kapusta, and K. A. Olive, Nucl.\ Phys.\ {\bf B348} (1991) 345.

\bibitem{baym76:a}
G. Baym and S. Chin, Phys.\ Lett.\ {\bf 62B} (1976) 241.

\bibitem{keister76:a}
B. D. Keister and L. S. Kisslinger, Phys.\ Lett.\ {\bf 64B} (1976) 117.

\bibitem{chap77:a}
G. Chapline and M. Nauenberg, Phys.\ Rev.\ D {\bf 16} (1977) 450.

\bibitem{fech78:a}
W. B. Fechner and P. C. Joss, Nature {\bf 274} (1978) 347.

\bibitem{bethe87:a}
H. A. Bethe, G. E. Brown, and J. Cooperstein, Nucl.\ Phys.\ {\bf A462} (1987)
  791.

\bibitem{serot87:a}
B. D. Serot and H. Uechi, Ann.\ Phys.\ (N. Y.)\ {\bf 179} (1987) 272.

\bibitem{arnett77:a}
W. D. Arnett and R. L. Bowers, Astrophys.\ J.\ Suppl. {\bf 33} (1977) 415.

\bibitem{taylor89:a}
J. H. Taylor and J. M. Weisberg, Astrophys.\ J.\ {\bf 345} (1989) 434.

\bibitem{thorsett93:a}
S. E. Thorsett, Z. Arzoumanian, M. M. McKinnon, and J. H. Taylor, Astrophys.\
  J.\ {\bf 405} (1993) L29.

\bibitem{rappaport83:a}
S. A. Rappaport and P. C. Joss, {\it Accrection Driven Stellar X-Ray Sources},
  ed. by W. H. G. Lewin and E. P. J. van den Heuvel, Cambridge University
  Press, 1983.

\bibitem{nagase89:a}
F. Nagase, Publ.\ Astron.\ Soc.\ Japan {\bf 41} (1989) 1.

\bibitem{glen90:ss}
N. K. Glendenning, {\it Nuclear and Particle Astrophysics}, Proc. Int. Summer
  School on the Structure of Hadrons and Hadronic Matter, Dronten, Netherlands,
  August 5-18, 1990, ed. by O. Scholten and J. H. Koch, World Scientific,
  Singapore, 1991, p. 275.

\bibitem{paradijs78:a}
J. Van Paradijs, Astrophys.\ J.\ {\bf 234} (1978) 609.

\bibitem{shapiro83:a}
S. L. Shapiro and S. A. Teukolsky, {\it Black Holes, White Dwarfs, and Neutron
  Stars}, John Wiley \& sons, N. Y., 1983.

\bibitem{fujimoto86:a}
M. Y. Fujimoto and R. E. Taam, Astrophys.\ J.\ {\bf 305} (1986) 246.

\bibitem{ruderman72:a}
M. Ruderman, Ann.\ Rev.\ Astr.\ Ap.\ {\bf 10} (1972) 427.

\bibitem{baym75:a}
G. Baym and C. Pethick, Ann.\ Rev.\ Nucl.\ Sci.\ {\bf 25} (1975) 27.

\bibitem{trimble70:a}
V. Trimble and M. Rees, Astrophys.\ Lett.\ {\bf 5} (1970) 93.

\bibitem{borner73:a}
G. Borner and J. M. Cohen, Astrophys.\ J.\ {\bf 185} (1973) 959.

\bibitem{liang86:a}
E. P. Liang, Astrophys.\ J.\ {\bf 304} (1986) 682.

\bibitem{brecher80:a}
K. Brecher and A. Burrows, Astrophys.\ J.\ {\bf 240} (1980) 642.

\bibitem{ramaty81:a}
R. Ramaty and P. Meszaros, Astrophys.\ J.\ {\bf 250} (1981) 384.

\bibitem{oppenheimer39}
J. R. Oppenheimer and G. M. Volkoff, Phys.\ Rev.\ {\bf 55} (1939) 374.

\bibitem{misner73:a}
Ch. W. Misner, K. S. Thorne, and J. A. Wheeler, {\it Gravitation}, W. H.
  Freeman and Company, San Francisco, 1973.

\bibitem{ruffini78:a}
R. Ruffini, in {\it Physics and Astrophysics of Neutron Stars and Black Holes}
  (North Holland, Amsterdam, 1978) p. 287.

\bibitem{brown94:a}
G. E. Brown and H. A. Bethe, Astrophys.\ J.\ {\bf 423} (1994) 659.

\bibitem{bethe95:a}
H. A. Bethe and G. E. Brown, Astrophys.\ J.\ {\bf 445} (1995) L129.

\bibitem{glen92:crust}
N. K. Glendenning and F. Weber, Astrophys.\ J.\ {\bf 400} (1992) 647.

\bibitem{lindblom86:a}
L. Lindblom, Astrophys.\ J.\ {\bf 303} (1986) 146.

\bibitem{lindblom92:a}
L. Lindblom, {\it Instabilities in Rotating Neutron Stars}, in The Structure
  and Evolution of Neutron Stars, Proceedings, ed. by D. Pines, R. Tamagaki,
  and S. Tsuruta, Addison-Wesley, 1992.

\bibitem{weber90:e}
F. Weber and N. K. Glendenning, Z.\ Phys.\ {\bf A339} (1991) 211.

\bibitem{chandrasekhar70:a}
S. Chandrasekhar, Phys.\ Rev.\ Lett.\ {\bf 24} (1970) 611.

\bibitem{friedman83:a}
J. L. Friedman, Phys.\ Rev.\ Lett.\ {\bf 51} (1983) 11.

\bibitem{lindblom77:a}
L. Lindblom and S. L. Detweiler, Astrophys.\ J.\ {\bf 211} (1977) 565.

\bibitem{detweiler75:a}
S. L. Detweiler, Astrophys.\ J.\ {\bf 197} (1975) 203.

\bibitem{lamb81:a}
H. Lamb, Proc.\ London\ Math.\ Soc. {\bf 13} (1881) 51.

\bibitem{cutler87:a}
C. Cutler and L. Lindblom, Astrophys.\ J.\ {\bf 314} (1987) 234.

\bibitem{ipser89:a}
J. R. Ipser and L Lindbolm, Phys.\ Rev.\ Lett.\ {\bf 62} (1989) 2777.

\bibitem{ipser90:a}
J. R. Ipser and L. Lindblom, Astrophys.\ J.\ {\bf 355} (1990) 226.

\bibitem{managan86:a}
R. A. Managan, Astrophys.\ J.\ {\bf 309} (1986) 598.

\bibitem{sawyer89:b}
R. F. Sawyer, Phys.\ Rev.\ D {\bf 39} (1989) 3804.

\bibitem{ipser91:a}
J. R. Ipser and L. Lindblom, Astrophys.\ J.\ {\bf 373} (1991) 213.

\bibitem{weber91:c}
F. Weber and N. K. Glendenning, {\it Impact of the Nuclear Equation of State on
  Models of Rotating Neutron Stars}, Proc. of the Int. Workshop on Unstable
  Nuclei in Astrophysics, Tokyo, Japan, June 7-8, 1991, ed. by S. Kubono and T.
  Kajino, World Scientific, Singapore, 1992, p. 307.

\bibitem{glen92:limit}
N. K. Glendenning, Phys.\ Rev.\ D {\bf 46} (1992) 4161.

\bibitem{glen93:drag}
N. K. Glendenning and F. Weber, Phys.\ Rev.\ D {\bf 50} (1994) 3836.

\bibitem{glen91:a}
N. K. Glendenning, Mod.\ Phys.\ Lett.\ {\bf A5} (1990) 2197.

\bibitem{weber92:a}
F. Weber and N. K. Glendenning, {\it Interpretation of Rapidly Rotating
  Pulsars}, Proceedings of the Second International Symposium on Nuclear
  Astrophysics, ``Nuclei in the Cosmos'', Karlsruhe, Germany, July 6--10, 1992,
  ed. by F. K{\"{a}}ppeler and K. Wisshak, IOP Publishing Ltd, Bristol, UK,
  1993, p. 399.

\bibitem{bodmer71:a}
A. R. Bodmer, Phys.\ Rev.\ D {\bf 4} (1971) 1601.

\bibitem{terazawa89:a}
H. Terazawa, INS-Report-338 (INS, Univ.\ of Tokyo, 1979); J.\ Phys.\ Soc.\
  Japan, {\bf 58} (1989) 3555; {\bf 58} (1989) 4388; {\bf 59} (1990) 1199.

\bibitem{madsen91:a}
J. Madsen and M. L. Olesen, Phys.\ Rev.\ D {\bf 43} (1991) 1069, ibid., {\bf
  44}, 566 (erratum).

\bibitem{caldwell91:a}
R. R. Caldwell and J. L. Friedman, Phys.\ Lett.\ {\bf 264B} (1991) 143.

\bibitem{baym71:a}
G. Baym, C. Pethick, and P. Sutherland, Astrophys.\ J.\ {\bf 170} (1971) 299.

\bibitem{pizzochero91:a}
P. Pizzochero, Phys.\ Rev.\ Lett.\ {\bf 66} (1991) 2425.

\bibitem{schaab95:a}
Ch. Schaab, F. Weber, M. K. Weigel, and N. K. Glendenning, Nucl.\ Phys.\ {\bf
  A605} (1996) 531.

\bibitem{weber93:b}
N. K. Glendenning, Ch. Kettner, and F. Weber, Astrophys.\ J.\ {\bf 450} (1995)
  253.

\bibitem{glen94:a}
N. K. Glendenning, Ch. Kettner, and F. Weber, Phys.\ Rev.\ Lett.\ {\bf 74}
  (1995) 3519.

\end{thebibliography}
\end{document}